\documentclass[apj,iop,numberedappendix,appendixfloats,twocolappendix,revtex4]{emulateapj}
\usepackage{apjfonts}

\usepackage{lscape,amsmath,graphicx,natbib,longtable,rotating}

\newcommand{\etal}{et al.}

\def\Ncen{N_{\rm cen}}
\def\Nsat{N_{\rm sat}}
\def\Mmin{M_{\rm min}}
\def\siglgM{\sigma_{\rm logM}}
\def\hinvMsun{h^{-1}M_\odot}
\def\hinvMpc{h^{-1}{\rm Mpc}}

\begin{document}

\title{Cross-Correlation of SDSS DR7 Quasars and DR10 BOSS Galaxies:
  The Weak Luminosity Dependence of Quasar Clustering at $z\sim0.5$}

\shorttitle{Quasar-Galaxy Cross Correlations in SDSS}


\shortauthors{SHEN ET AL.}
\author{Yue Shen\altaffilmark{1,2,3},
Cameron K. McBride\altaffilmark{1}, Martin White\altaffilmark{4,5,6}, Zheng
Zheng\altaffilmark{7}, Adam D. Myers\altaffilmark{8}, Hong
Guo\altaffilmark{10}, Jessica A. Kirkpatrick\altaffilmark{4,5}, Nikhil
Padmanabhan\altaffilmark{9}, John K. Parejko\altaffilmark{9}, Nicholas P.
Ross\altaffilmark{4}, David J. Schlegel\altaffilmark{4}, Donald P.
Schneider\altaffilmark{11,12}, Alina Streblyanska\altaffilmark{13,14}, Molly
E.~C. Swanson\altaffilmark{1}, Idit Zehavi\altaffilmark{10}, Kaike
Pan\altaffilmark{15}, Dmitry Bizyaev\altaffilmark{15}, Howard
Brewington\altaffilmark{15}, Garrett Ebelke\altaffilmark{15}, Viktor
Malanushenko\altaffilmark{15}, Elena Malanushenko\altaffilmark{15}, Daniel
Oravetz\altaffilmark{15}, Audrey Simmons\altaffilmark{15}, Stephanie
Snedden\altaffilmark{15}}


\altaffiltext{1}{Harvard-Smithsonian Center for Astrophysics, 60 Garden
Street, MS-51, Cambridge, MA 02138, USA}

\altaffiltext{2}{Carnegie Observatories, 813 Santa Barbara Street, Pasadena,
CA 91101, USA}

\altaffiltext{3}{Hubble Fellow}

\altaffiltext{4}{Lawrence Berkeley National Laboratory, One Cyclotron Road,
Berkeley, CA 94720, USA}

\altaffiltext{5}{Department of Physics, University of California, Berkeley,
CA 94720, USA}

\altaffiltext{6}{Department of Astronomy, University of California, Berkeley,
CA 94720, USA}

\altaffiltext{7}{Department of Physics and Astronomy, University of Utah,
Salt Lake City, UT 84112, USA}

\altaffiltext{8}{Department of Physics and Astronomy, University of Wyoming,
Laramie, WY 82071, USA}

\altaffiltext{9}{Yale Center for Astronomy and Astrophysics, Yale University,
New Haven, CT, 06520, USA}

\altaffiltext{10}{Department of Astronomy, Case Western Reserve University,
OH 44106, USA}

\altaffiltext{11}{Department of Astronomy and Astrophysics, The Pennsylvania
State University, University Park, PA 16802, USA}

\altaffiltext{12}{Institute for Gravitation and the Cosmos, The Pennsylvania
State University, University Park, PA 16802, USA}

\altaffiltext{13}{Instituto de Astrofisica de Canarias (IAC), E-38200 La
Laguna, Tenerife, Spain}

\altaffiltext{14}{Dept. Astrofisica, Universidad de La Laguna (ULL), E-38206
La Laguna, Tenerife, Spain}

\altaffiltext{15}{Apache Point Observatory, P.O. Box 59, Sunspot, NM
88349-0059, USA}

\begin{abstract}

We present the measurement of the two-point cross-correlation function (CCF)
of $8,198$ Sloan Digital Sky Survey (SDSS) Data Release 7 (DR7) quasars and
$349,608$ DR10 CMASS galaxies from the Baryonic Oscillation Spectroscopic
Survey (BOSS) at redshift $\bar{z}\sim 0.5$ ($0.3<z<0.9$). The
cross-correlation function can be reasonably well fit by a power-law model
$\xi_{QG}(r)=(r/r_0)^{-\gamma}$ on projected scales of $r_p=2-25\,h^{-1}{\rm
Mpc}$ with $r_0=6.61\pm0.25\,h^{-1}$Mpc and $\gamma=1.69\pm0.07$. We estimate
a quasar linear bias of $b_Q=1.38\pm0.10$ at $\langle z\rangle=0.53$ from the
CCF measurements. This linear bias corresponds to a characteristic host halo
mass of $\sim 4\times 10^{12}\,h^{-1}M_\odot$, compared to $\sim
10^{13}\,h^{-1}M_\odot$ characteristic host halo mass for CMASS galaxies.
Based on the clustering measurements, most quasars at $\bar{z}\sim 0.5$ are
not the descendants of their higher luminosity counterparts at higher
redshift, which would have evolved into more massive and more biased systems
at low redshift. We divide the quasar sample in luminosity and constrain the
luminosity dependence of quasar bias to be $db_{Q}/d\log L=0.20\pm0.34$ or
$0.11\pm 0.32$ (depending on different luminosity divisions) for quasar
luminosities $-23.5>M_i(z=2)>-25.5$, implying a weak luminosity dependence of
quasar clustering for the bright end of the quasar population at
$\bar{z}\sim0.5$. We compare our measurements with theoretical predictions,
Halo Occupation Distribution (HOD) models and mock catalogs. These
comparisons suggest quasars reside in a broad range of host halos, and the
host halo mass distributions significantly overlap with each other for
quasars at different luminosities, implying a poor correlation between halo
mass and instantaneous quasar luminosity. We also find that the quasar HOD
parameterization is largely degenerate such that different HODs can reproduce
the CCF equally well, but with different outcomes such as the satellite
fraction and host halo mass distribution. These results highlight the
limitations and ambiguities in modeling the distribution of quasars with the
standard HOD approach and the need for additional information in populating
quasars in dark matter halos with HOD.
\end{abstract}

\section{introduction}\label{sec:intro}

Quasars are powered by mass accretion onto supermassive black holes (SMBHs)
at the center of massive galaxies. Like galaxies, quasars are luminous
tracers of the underlying dark matter, and can be used to map the large-scale
structure of the Universe. Over the past decade, quasar clustering has been
measured for large statistical samples drawn from dedicated surveys, most
notably the Sloan Digital Sky Survey \citep[SDSS,][]{SDSS} and the 2dF QSO
Redshift Survey \citep[2QZ,][]{Croom_etal_2004}. Building on earlier studies
on small and heterogenous samples \citep[e.g.,][]{Shaver_1984}, the
auto-correlation function of quasars has been measured with unprecedented
precision for a wide redshift range (from $z\sim 0.4$ to $z\sim 4$) and a
variety of quasar properties
\citep[e.g.,][]{PMN_2004,Croom_etal_2005,Porciani_Norberg_2006,Myers_etal_2006,Myers_etal_2007a,Myers_etal_2007b,
Shen_etal_2007b,Shen_etal_2008a,Shen_etal_2009a,da_Angela_2005,
da_Angela_2008,Ross_etal_2009,Ivashchenko_etal_2010,White_etal_2012}, and has
been extended to the small-scale regime \citep[$\lesssim 1\,h^{-1}{\rm Mpc}$,
e.g.,][]{Hennawi_etal_2006,Myers_etal_2008,Shen_etal_2010,Kayo_Oguri_2012}.
The clustering measurements have also been performed for Active Galactic
Nuclei (AGNs) selected at non-optical wavelengths
\citep[e.g.,][]{Wake_etal_2008,Gilli_etal_2009,Coil_etal_2009,Hickox_etal_2009,Hickox_etal_2011,
Donoso_etal_2010,Cappelluti_etal_2010,Krumpe_etal_2010,
Krumpe_etal_2012,Miyaji_etal_2011,Allevato_etal_2011}. These quasar/AGN
clustering measurements revealed that quasars live in massive ($\sim
10^{12}-10^{13}\, h^{-1}M_\odot$) dark matter halos, and constraints on the
duty cycle of quasar activity can be inferred from the relative abundance of
quasars and their host halos
\citep[e.g.,][]{Cole_Kaiser_1989,Martini_Weinberg_2001,Haiman_Hui_2001}.

With quasar samples increasing in size, several attempts have been made to
measure quasar clustering as a function of quasar luminosity. More massive
halos are formed in rarer peaks of the density fluctuation field and are more
strongly clustered
\citep[e.g.,][]{BBKS,Cole_Kaiser_1989,Mo_White_1996,Sheth_Mo_Tormen_2001}.
Galaxy clustering shows a strong dependence on luminosity
\citep[e.g.,][]{Norberg_etal_2001,Zehavi_etal_2005,Zehavi_etal_2011,Coil_etal_2006,Coupon_etal_2012},
indicating a good correlation between host halo mass and galaxy luminosity.
On the other hand, quasar clustering studies to date have failed to detect a
strong luminosity dependence
\citep[e.g.,][]{Adelberger_Steidel_2005,Porciani_Norberg_2006,Myers_etal_2007a,da_Angela_2008,White_etal_2012},
although \citet{Shen_etal_2009a} reported a 2$\sigma$ detection for the most
luminous quasars in SDSS Data Release 5 (DR5) at $\langle z\rangle\sim 1.5$.

A weak dependence of quasar clustering on luminosity is expected if quasar
luminosity is not tightly correlated with halo mass. Scatter between the
instantaneous quasar luminosity and host halo mass dilutes any luminosity
dependence of the clustering. Several semi-analytical cosmological quasar
models have been constructed to make predictions broadly consistent with
current constraints on the luminosity dependence of quasar clustering
\citep[e.g.,][for recent
work]{Lidz_etal_2006,Shen_2009,Shankar_etal_2010,CW13}; more sophisticated
approaches with dark matter-only simulations$+$semi-analytical galaxy
formation models
\citep[e.g.,][]{Bonoli_etal_2009,Fanidakis_etal_2012,Hirschmann_etal_2012},
or with fully hydrodynamic cosmological simulations
\citep[e.g.,][]{Thacker_etal_2009,Degraf_etal_2011,Chatterjee_etal_2012} are
underway. Precise measurements of the luminosity dependence of quasar
clustering are important in quantifying the scatter between quasar luminosity
and host halo mass \citep[e.g.,][]{White_etal_2008,Shankar_etal_2010}, which
can in turn provide useful constraints on the correlation between black hole
mass and halo mass, and on quasar light curve models
\citep[e.g.,][]{Yu_Lu_2004,Yu_Lu_2008,Hopkins_etal_2005a,Hopkins_etal_2008a,Shen_2009,Croton_2009,Cao_2010,Shanks_etal_2011}.

The sparseness of quasars makes the measurements of the luminosity dependence
of quasar clustering a nontrivial task. Fine bins in luminosity and redshift,
while breaking the $L-z$ degeneracy, lead to very noisy clustering
measurements \citep[e.g.,][]{da_Angela_2008}, hampering the detection of a
possible luminosity dependence. \citet{Shen_etal_2009a} used a flux-limited
quasar sample covering a wide redshift range ($0.4<z<2.5$) in order to
increase the statistics, but the resulting luminosity subsamples are mixtures
over a range of quasar luminosity and redshift.

One approach to mitigate such poor statistics is to cross-correlate the
quasar sample with a much larger, galaxy sample. On large scales, where
linear bias applies, the cross-correlation function is determined by the
auto-correlation functions of both sets of tracers. Using the
cross-correlation technique, one can obtain a much better measurement of
quasar clustering by boosting the pair counts, suppressing the shot noise
from the small number of pairs in quasar auto-correlation measurements. In
addition, the small-scale cross correlation between galaxies and quasars
constrains the occupation of galaxies in quasar-hosting halos, and may hint
on the triggering mechanism of quasar activity.

There have been a number of studies on the cross correlation between galaxies
and different types of quasars and Active Galactic Nuclei (AGN), i.e.,
optical-selected quasars, X-ray-, radio- and infrared-selected (type 1 and
type 2) AGNs
\cite[e.g.,][]{Adelberger_Steidel_2005,Li_etal_2006,Coil_etal_2007,Coil_etal_2009,Wake_etal_2008,
Padmanabhan_etal_2009,Donoso_etal_2010,Krumpe_etal_2010,
Krumpe_etal_2012,Miyaji_etal_2011,Hickox_etal_2009,Hickox_etal_2011}. These
studies generally found weak or no luminosity dependence of the large-scale
quasar bias, although these measurements can be improved upon using larger
samples.

Here we use the Tenth Data Release (DR10), ``CMASS'', galaxy sample
\citep{White_etal_2011,Anderson_etal_2012,Sanchez_etal_2012} from the Baryon
Oscillation Spectroscopic Survey \citep[BOSS;][]{SchWhiEis09,Dawson_12} in
SDSS III \citep{Eisenstein_etal_2011} and the DR7 \citep{Abazajian_etal_2009}
spectroscopic quasar sample from SDSS I/II \citep[][]{Schneider_etal_2010} to
measure the cross correlation function of galaxies and quasars at $0.3<z<0.9$
($\langle z\rangle\sim 0.53$). These samples represent the largest and most
homogeneous spectroscopic samples to date for such cross correlation
analyses, and enable us to derive one of the most stringent constraints on
the luminosity dependence of large-scale quasar clustering in this redshift
range. It also provides important clues on how galaxies and quasars occupy
the same dark matter halos as functions of galaxy and quasar properties, thus
shedding light on the assembly process of quasars and their immediate
environment.

In this study we focus on the luminosity dependence of quasar linear bias at
$z\sim 0.5$, although we also briefly touch on the occupation of quasars
within dark matter halos. More detailed modeling and discussions on the other
interesting aspects of quasar-galaxy cross-correlation will be reported in
future work. This paper is organized as follows: \S\ref{sec:sample} describes
the quasar and galaxy samples used; the cross correlation function
measurements are presented in \S\ref{sec:cf}; we present a detailed
discussion on our results in terms of comparisons to theoretical quasar
models (\S\ref{sec:disc1}), Halo Occupation Distribution (HOD) modeling
(\S\ref{sec:disc2}), and mock catalog based interpretation
(\S\ref{sec:disc3}); we conclude in \S\ref{sec:con}. In the Appendix we
present systematic checks of our correlation function measurements.
Throughout the paper we adopt a flat $\Lambda$CDM cosmology with
$\Omega_{\Lambda}=0.726$, $h=0.7$, $\Omega_{b}=0.0457$, $\sigma_8=0.8$ and
$n_s=0.95$ \citep[e.g.,][]{Komatsu_etal_2011}. All errors quoted are
$1\sigma$ statistical only, unless otherwise specified. Quasar luminosities
are quoted in terms of $M_i(z=2)$, the absolute $i$ band magnitude normalized
at $z=2$ \citep{Richards_etal_2006a}.

\section{The Data}\label{sec:sample}

\begin{figure}
 \centering
 \includegraphics[width=0.9\linewidth]{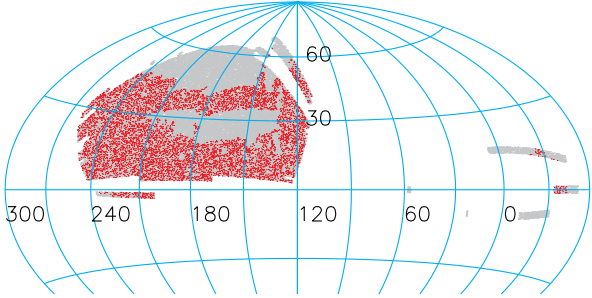}
 \caption{Aitoff projection of the sky coverage of the cross-correlation samples. The gray region shows the
 entire SDSS DR7 uniform quasar sample footprint, while the red region shows the current
 overlap with the DR10 BOSS CMASS galaxy sample.
 }
 \label{fig:sky}
\end{figure}

\begin{figure}
 \centering
 \includegraphics[width=0.9\linewidth]{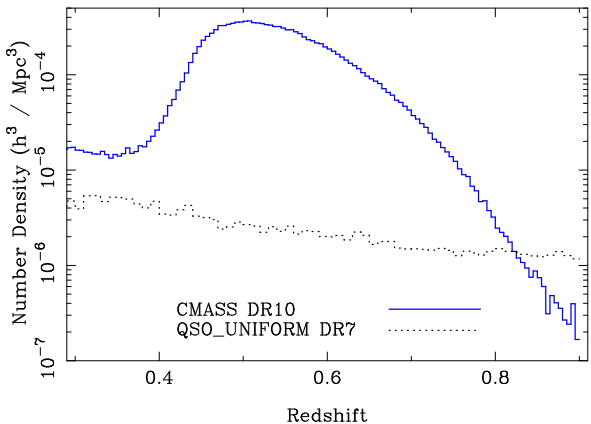}
 \caption{Number density as a function of redshift for the DR7 uniform quasar and DR10 CMASS galaxy
 samples. We have limited both samples within $0.3<z<0.9$. }
 \label{fig:numden}
\end{figure}

\begin{figure*}
 \centering
 \includegraphics[width=0.45\textwidth]{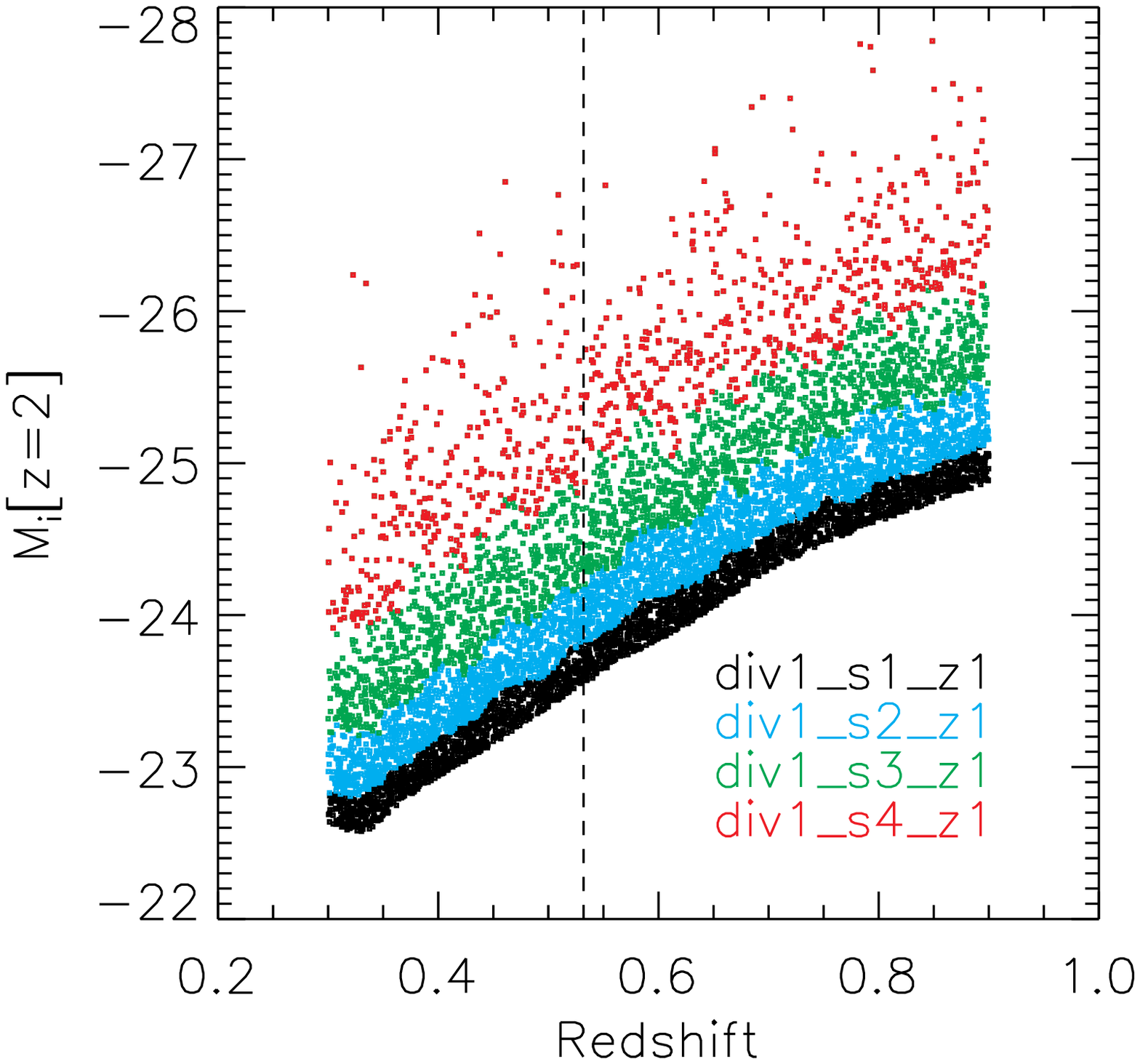}
 \includegraphics[width=0.45\textwidth]{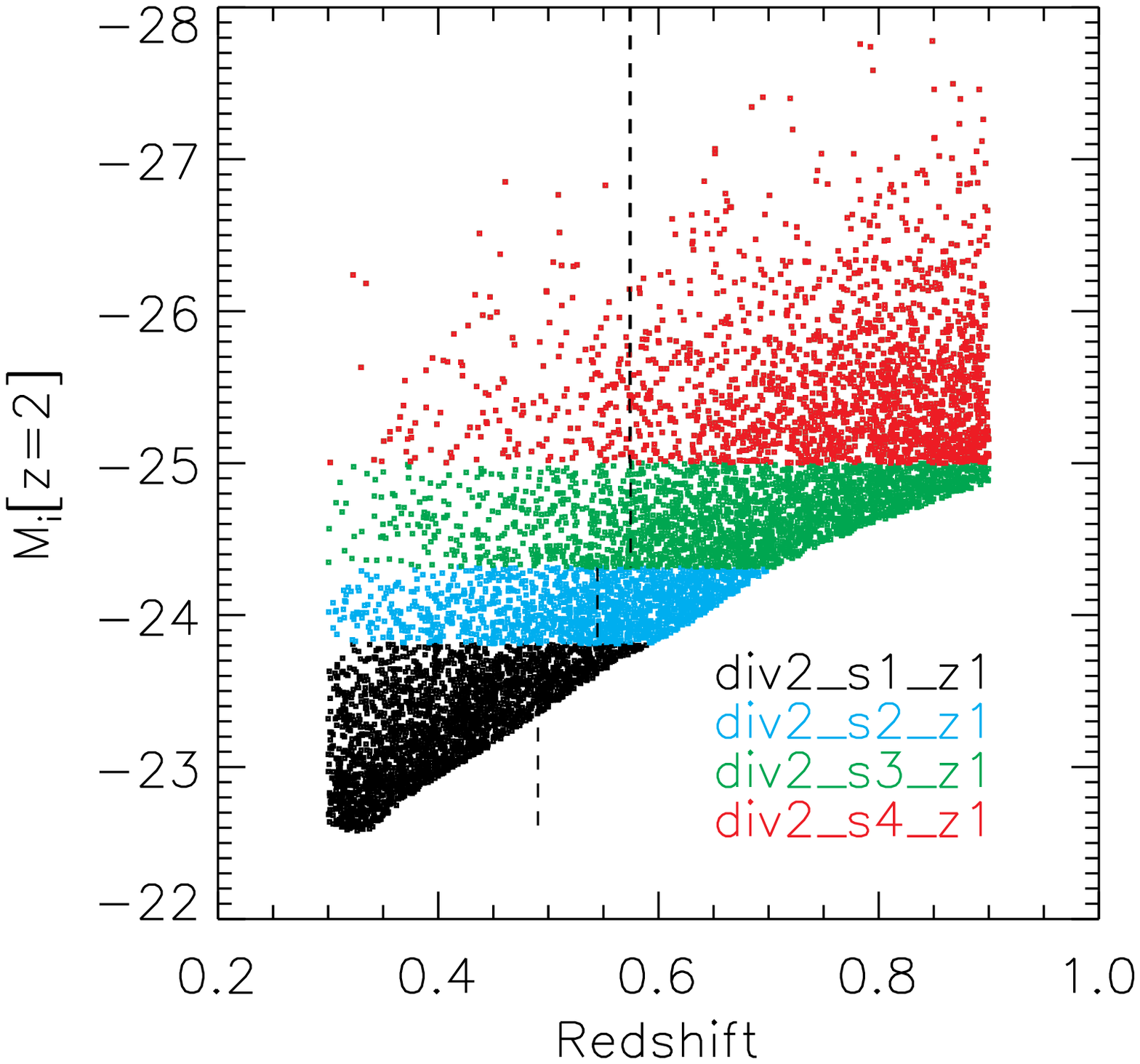}
 \includegraphics[width=0.45\textwidth]{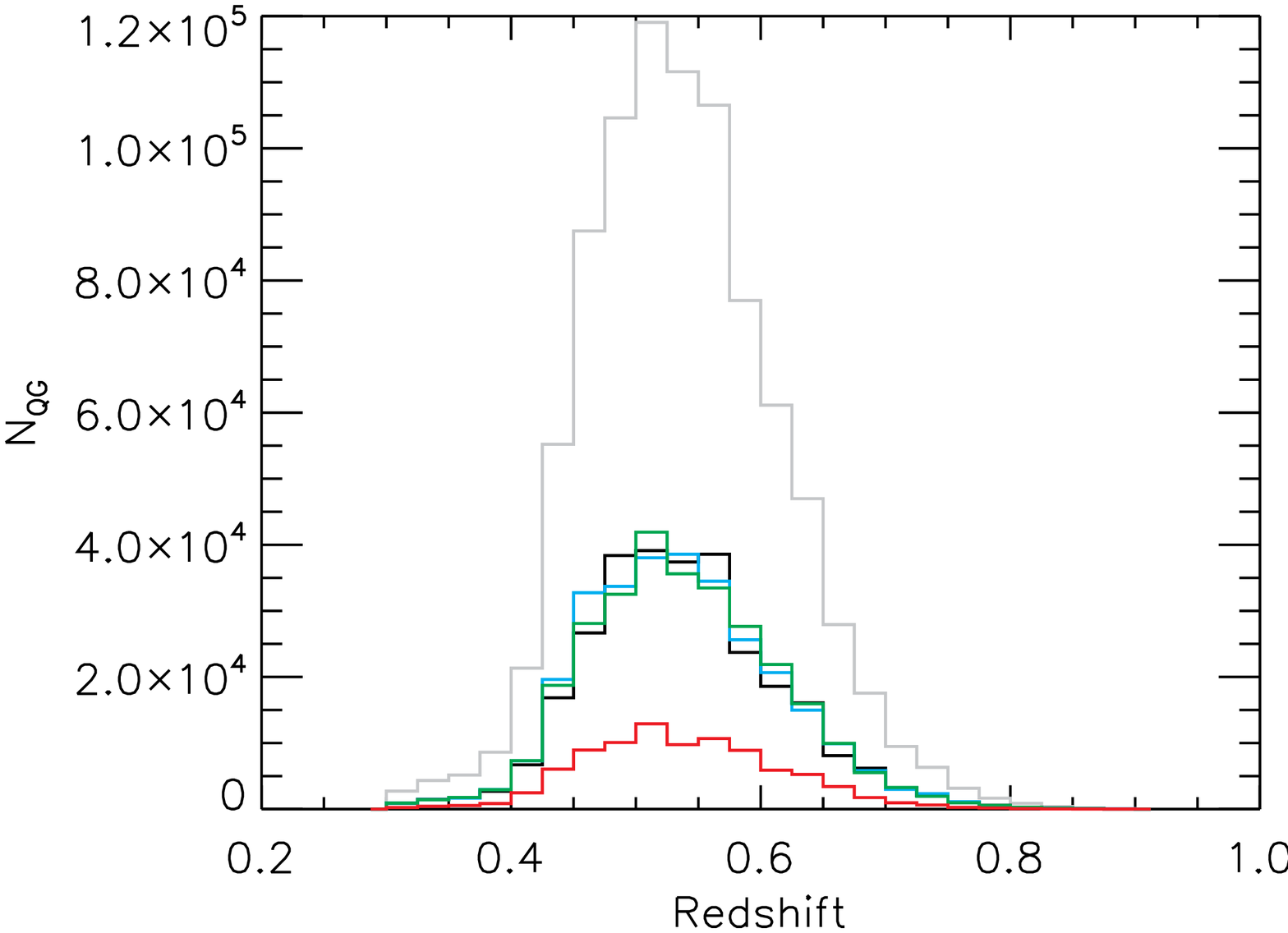}
 \includegraphics[width=0.45\textwidth]{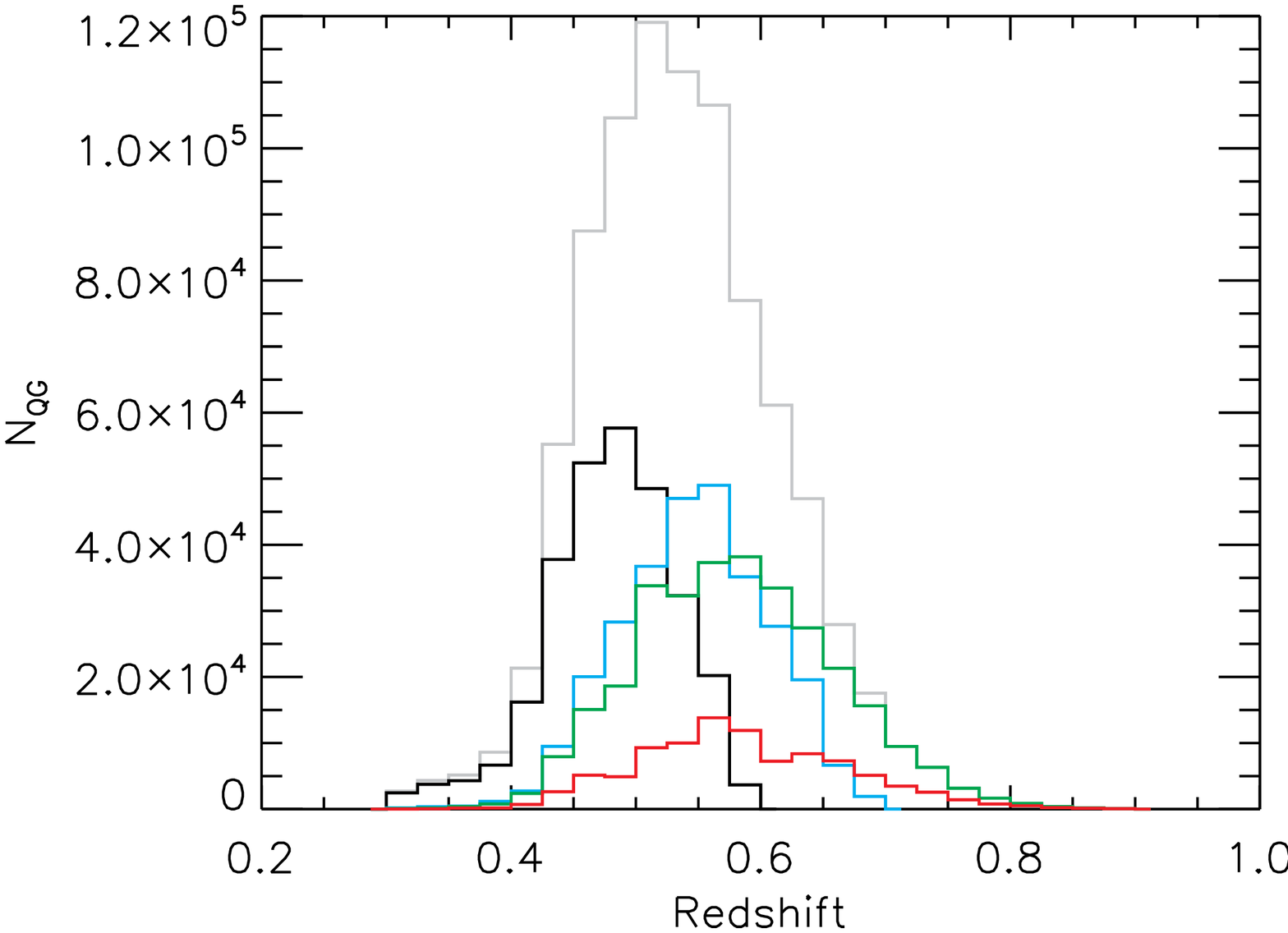}
    \caption{Subsamples of quasars divided by quasar luminosity. The detailed sample definition is
    described in \S\ref{sec:subsample} and summarized in Table \ref{tab:summary}. The
    top panels show the distribution in the quasar luminosity-redshift plane,
    with different colors for the four different luminosity subsamples. Note that the red points overlap
    with the green points, i.e., the most luminous subsample is a subset of a less luminous subsample. The vertical dashed lines further split each
    luminosity subsample by the cross-pair-weighted
    median redshift. The bottom panels show the
    cross pair-weighted (with $QG$ pair separations $r_p<50\,h^{-1}$Mpc and $\pi<70\,h^{-1}$Mpc) redshift distribution of
    quasars in each subsample (with
    the gray lines showing that for the full sample). The left and right
    columns are for Division 1 and Division 2 in terms of quasar luminosity, respectively.
    }
    \label{fig:mi_z_dist}
\end{figure*}

\begin{table*}
\caption{Summary of Quasar Subsamples. $N^*_{\rm QG}$ is the total number of
quasar-galaxy pairs with $r_p<50\,h^{-1}{\rm Mpc}$ and $\pi<70\,h^{-1}{\rm
Mpc}$ in a given cross-correlation sample. The median redshift and magnitude
are the pair-count (with $r_p<50\,h^{-1}{\rm Mpc}$ and $\pi<70\,h^{-1}{\rm
Mpc}$) weighted median values of quasars. The last four columns list the
best-fit power-law model correlation length of the CCF (with fixed slope
$\gamma=1.7$), the galaxy linear bias, the linear bias of the
cross-correlation sample fitted with the full covariance matrix and with
diagonal elements of the covariance matrix. See \S\ref{sec:cf} for details on
subsamples and the estimation of correlation lengths and linear biases.}
\scalebox{0.9}{
\begin{tabular}{llccccccccccccc}
\hline
\# & Sample & $N_{\rm Q}$ & $N_{\rm G}$ &
$N_{\rm QG}^*$ & $z_{\rm min}$ & $z_{\rm max}$ & $M_{i,\rm{min}}$ &
$M_{i,\rm{max}}$ & $\langle z\rangle$ & $\langle M_{i}\rangle$ &
$r_0(\gamma=1.7)$ & $b_{\rm G}$ & $b_{\rm QG}$ & $b_{\rm QG}^{\rm diag}$ \\
\hline
0  & Full           &    8198  &  349608  &  879352 & 0.3000 & 0.8999 &
$-28.693$ & $-22.576$ & 0.532 & $-24.055$ & $6.614_{-0.240}^{+0.234}$ &
$2.10\pm 0.02$ & $1.70_{-0.06}^{+0.06}$ & $1.70\pm 0.04$ \\
\hline
1  & div1\_s1\_z1   &    2726  &  349608  &  293098 & 0.3003 & 0.8998 &
$-25.115$ & $-22.576$ & 0.533 & $-23.675$ & $6.682_{-0.433}^{+0.414}$ &
$2.11\pm 0.02$ & $1.69_{-0.11}^{+0.11}$ & $1.72\pm 0.07$ \\
2  & div1\_s1\_z2   &    1075  &  155888  &  134524 & 0.3003 & 0.5320 &
$-23.819$ & $-22.576$ & 0.481 & $-23.440$ & $6.390_{-0.654}^{+0.610}$ &
$2.03\pm 0.04$ & $1.44_{-0.17}^{+0.16}$ & $1.42\pm 0.10$ \\
3  & div1\_s1\_z3   &    1651  &  193720  &  135256 & 0.5321 & 0.8998 &
$-25.115$ & $-23.570$ & 0.589 & $-23.942$ & $6.966_{-0.535}^{+0.508}$ &
$2.15\pm 0.03$ & $1.90_{-0.16}^{+0.15}$ & $2.01\pm 0.09$ \\
4  & div1\_s2\_z1   &    2738  &  349608  &  293640 & 0.3002 & 0.8999 &
$-25.541$ & $-22.808$ & 0.531 & $-24.000$ & $6.841_{-0.327}^{+0.316}$ &
$2.10\pm 0.02$ & $1.69_{-0.08}^{+0.08}$ & $1.69\pm 0.06$ \\
5  & div1\_s2\_z2   &    1068  &  155888  &  137808 & 0.3002 & 0.5319 &
$-24.171$ & $-22.808$ & 0.480 & $-23.726$ & $6.899_{-0.463}^{+0.442}$ &
$2.03\pm 0.04$ & $1.69_{-0.11}^{+0.10}$ & $1.68\pm 0.08$ \\
6  & div1\_s2\_z3   &    1670  &  193720  &  133358 & 0.5322 & 0.8999 &
$-25.541$ & $-23.838$ & 0.591 & $-24.294$ & $6.856_{-0.450}^{+0.431}$ &
$2.15\pm 0.03$ & $1.73_{-0.12}^{+0.11}$ & $1.72\pm 0.09$ \\
7  & div1\_s3\_z1   &    2734  &  349608  &  292614 & 0.3000 & 0.8993 &
$-28.693$ & $-23.208$ & 0.533 & $-24.727$ & $6.277_{-0.358}^{+0.344}$ &
$2.11\pm 0.02$ & $1.70_{-0.10}^{+0.09}$ & $1.67\pm 0.07$ \\
8  & div1\_s3\_z2   &    1069  &  155888  &  135812 & 0.3000 & 0.5319 &
$-26.851$ & $-23.208$ & 0.481 & $-24.395$ & $6.823_{-0.607}^{+0.571}$ &
$2.03\pm 0.04$ & $1.78_{-0.15}^{+0.14}$ & $1.79\pm 0.09$ \\
9  & div1\_s3\_z3   &    1665  &  193720  &  133933 & 0.5327 & 0.8993 &
$-28.693$ & $-24.204$ & 0.591 & $-24.991$ & $5.303_{-0.573}^{+0.533}$ &
$2.15\pm 0.03$ & $1.58_{-0.14}^{+0.13}$ & $1.52\pm 0.10$ \\
10 & div1\_s4\_z1   &     837  &  349608  &   91081 & 0.3004 & 0.8993 &
$-28.693$ & $-23.915$ & 0.533 & $-25.406$ & $6.804_{-0.429}^{+0.411}$ &
$2.11\pm 0.02$ & $1.79_{-0.13}^{+0.12}$ & $1.78\pm 0.10$ \\
11 & div1\_s4\_z2   &     321  &  155888  &   41766 & 0.3004 & 0.5303 &
$-26.851$ & $-23.915$ & 0.482 & $-25.043$ & $5.404_{-1.075}^{+0.942}$ &
$2.03\pm 0.04$ & $1.93_{-0.20}^{+0.18}$ & $1.88\pm 0.15$ \\
12 & div1\_s4\_z3   &     516  &  193720  &   42015 & 0.5329 & 0.8993 &
$-28.693$ & $-24.876$ & 0.592 & $-25.622$ & $5.634_{-0.942}^{+0.842}$ &
$2.15\pm 0.03$ & $1.39_{-0.28}^{+0.23}$ & $1.42\pm 0.17$ \\
\hline
13 & div2\_s1\_z1   &    2397  &  249546  &  283766 & 0.3000 & 0.5889 &
$-23.812$ & $-22.576$ & 0.484 & $-23.564$ & $6.861_{-0.460}^{+0.439}$ &
$2.05\pm 0.03$ & $1.67_{-0.11}^{+0.10}$ & $1.63\pm 0.07$ \\
14 & div2\_s1\_z2   &    1995  &   78593  &  136423 & 0.3000 & 0.4906 &
$-23.810$ & $-22.576$ & 0.448 & $-23.420$ & $6.797_{-0.655}^{+0.614}$ &
$2.14\pm 0.05$ & $1.55_{-0.16}^{+0.14}$ & $1.52\pm 0.10$ \\
15 & div2\_s1\_z3   &     402  &  170953  &  112867 & 0.4907 & 0.5889 &
$-23.812$ & $-23.369$ & 0.524 & $-23.659$ & $6.429_{-0.689}^{+0.641}$ &
$2.06\pm 0.04$ & $1.69_{-0.19}^{+0.17}$ & $1.78\pm 0.10$ \\
16 & div2\_s2\_z1   &    1443  &  335123  &  286117 & 0.3005 & 0.6980 &
$-24.315$ & $-23.812$ & 0.547 & $-24.040$ & $6.988_{-0.379}^{+0.365}$ &
$2.11\pm 0.02$ & $1.69_{-0.11}^{+0.10}$ & $1.69\pm 0.07$ \\
17 & div2\_s2\_z2   &     628  &  178865  &  123829 & 0.3005 & 0.5446 &
$-24.315$ & $-23.812$ & 0.499 & $-24.018$ & $5.744_{-0.576}^{+0.538}$ &
$2.05\pm 0.03$ & $1.42_{-0.13}^{+0.12}$ & $1.44\pm 0.10$ \\
18 & div2\_s2\_z3   &     815  &  156258  &  132738 & 0.5447 & 0.6980 &
$-24.315$ & $-23.813$ & 0.592 & $-24.066$ & $7.150_{-0.499}^{+0.475}$ &
$2.12\pm 0.03$ & $1.77_{-0.18}^{+0.16}$ & $1.84\pm 0.10$ \\
19 & div2\_s3\_z1   &    4358  &  349608  &  306945 & 0.3004 & 0.8999 &
$-28.693$ & $-24.315$ & 0.578 & $-24.741$ & $5.923_{-0.312}^{+0.301}$ &
$2.15\pm 0.02$ & $1.75_{-0.09}^{+0.09}$ & $1.74\pm 0.06$ \\
20 & div2\_s3\_z2   &     624  &  229499  &  138601 & 0.3004 & 0.5747 &
$-26.851$ & $-24.315$ & 0.518 & $-24.740$ & $6.108_{-0.487}^{+0.461}$ &
$2.09\pm 0.03$ & $1.88_{-0.13}^{+0.13}$ & $1.88\pm 0.09$ \\
21 & div2\_s3\_z3   &    3734  &  120109  &  143922 & 0.5748 & 0.8999 &
$-28.693$ & $-24.316$ & 0.637 & $-24.744$ & $6.259_{-0.371}^{+0.356}$ &
$2.19\pm 0.04$ & $1.77_{-0.09}^{+0.09}$ & $1.69\pm 0.07$ \\
22 & div2\_s4\_z1   &    1966  &  349608  &   95949 & 0.3019 & 0.8999 &
$-28.693$ & $-25.000$ & 0.579 & $-25.417$ & $6.030_{-0.546}^{+0.513}$ &
$2.15\pm 0.02$ & $1.75_{-0.13}^{+0.12}$ & $1.70\pm 0.11$ \\
23 & div2\_s4\_z2   &     188  &  228104  &   42244 & 0.3019 & 0.5738 &
$-26.851$ & $-25.003$ & 0.521 & $-25.406$ & $5.936_{-0.876}^{+0.794}$ &
$2.09\pm 0.03$ & $1.74_{-0.23}^{+0.20}$ & $1.69\pm 0.17$ \\
24 & div2\_s4\_z3   &    1778  &  121504  &   45791 & 0.5745 & 0.8999 &
$-28.693$ & $-25.000$ & 0.644 & $-25.419$ & $6.477_{-0.719}^{+0.667}$ &
$2.20\pm 0.04$ & $1.90_{-0.17}^{+0.16}$ & $1.84\pm 0.13$ \\
\hline
\end{tabular}}\label{tab:summary}
\end{table*}

%

The SDSS I/II uses a dedicated 2.5-m wide-field telescope
\citep{Gunn_etal_2006} with a drift-scan camera with 30 $2048 \times 2048$
CCDs \citep{Gunn_etal_1998} to image the sky in five broad bands
\citep[$u\,g\,r\,i\,z$;][]{Fukugita_etal_1996}. The imaging data are taken on
dark photometric nights of good seeing \citep{Hogg_etal_2001}, are calibrated
photometrically \citep{Smith_etal_2002, Ivezic_etal_2004, Tucker_etal_2006}
and astrometrically \citep{Pier_etal_2003}, and object parameters are
measured \citep{Lupton_etal_2001}. Quasar candidates
\citep{Richards_etal_2002a} for follow-up spectroscopy are selected from the
imaging data using their colors, and are arranged in spectroscopic plates
\citep{Blanton_etal_2003} to be observed with a pair of fiber-fed double
spectrographs \citep{Smee_etal_2012}. The final (DR7) quasar catalog from
SDSS I/II was presented in \citet{Schneider_etal_2010}.

The BOSS survey is an ongoing program within SDSS III
\citep{Eisenstein_etal_2011}, which is obtaining spectra for massive galaxy
and quasar targets selected using photometry from SDSS I/II and new imaging
data in the South Galactic Cap (SGC) in SDSS III. Targets are observed with
an upgraded version of the multi-object fiber spectrographs for SDSS I/II
\citep{Smee_etal_2012}. The BOSS spectra are reduced and classified by an
automatic pipeline described in \citet{Bolton_etal_2012}, and the first
public data release of BOSS spectra is Data Release 9 (DR9)
\citep{Ahn_etal_2012}. In this work we use the unpublished Data Release 10
(DR10) for our galaxy sample, which contains BOSS spectra taken through July
2012, and surpasses the DR9 samples.

\subsection{Sample Construction}

We use the subset of quasars in the SDSS DR7 quasar catalog
\citep{Schneider_etal_2010}, with \texttt{UNIFORM\_TARGET}$=1$ in the
value-added catalog of \citet{Shen_etal_2011}. These quasars were uniformly
targeted using the final quasar target selection algorithm
\citep{Richards_etal_2002a} implemented in SDSS I/II, and constitute a
statistical sample suitable for clustering studies
\citep[e.g.,][]{Shen_etal_2007b,Shen_etal_2009a,Ross_etal_2009}. For the
redshift range of interest here ($z<1$), this quasar sample is flux limited
to $i=19.1$. The sky coverage of this uniform quasar sample is 6248 deg$^2$.

Two main galaxy samples are targeted in BOSS, with separate color and
magnitude cuts: the CMASS sample at $\langle z\rangle\sim 0.55$, and the LOWZ
sample at $z\lesssim 0.4$. We choose the CMASS sample as our galaxy sample,
as it has a larger redshift overlap with our quasar sample. The total DR10
BOSS CMASS galaxy sample contains over $560\,k$ galaxies, which is
approximately one half of the final BOSS CMASS galaxy sample.

Since the CMASS galaxy sample has a narrow redshift distribution that peaks
around $z\sim 0.55$ and drops rapidly towards both ends, we have imposed a
redshift cut, $0.3<z<0.9$, to both the CMASS sample and the quasar sample.
Fig.\ \ref{fig:sky} shows the overlap between the CMASS galaxy sample and the
DR7 uniform quasar sample used in the current study, with a sky area of
$4122\,{\rm deg}^2$. Fig.\ \ref{fig:numden} shows the redshift distributions
of our final CMASS sample and quasar sample for subsequent cross-correlation
analysis, with 349,608 galaxies and 8,198 quasars in total.

\subsection{Quasar Luminosity Subsamples}\label{sec:subsample}

Since our primary goal is to investigate the luminosity dependence of quasar
clustering, we divide our quasar sample into different subsamples by quasar
luminosity.

The redshift distributions of the quasars and CMASS galaxies (e.g., Fig.\
\ref{fig:numden}) suggest that most of the pair contribution comes from a
rather narrow redshift range around $z\sim 0.5$. Thus any redshift-dependent
clustering is expected to be small. Nevertheless, we consider quasar
subsamples divided by redshift-varying luminosity boundaries (Division 1), as
well as by constant luminosity cuts (Division 2), as shown in Fig.\
\ref{fig:mi_z_dist}. Division 1 enforces all subsamples to have the same
redshift distribution, but the subsamples will overlap with each other in
luminosity. Division 2 ensures there is no luminosity overlap in each
subsample, but the effective redshift is slightly different for each
subsample. We further split these luminosity subsamples by the pair-weighted
quasar median redshift in each bin to create $L-z$ subsamples, to investigate
possible redshift evolution. Table \ref{tab:summary} summarizes the
luminosity and redshift boundaries and properties of these quasar subsamples.
These redshift and luminosity boundaries were chosen to yield comparable pair
counts for cross-correlation subsamples, except for the most luminous
subsamples (div1\_s4\_* and div2\_s4\_*).

We assign the effective luminosity and redshift to each quasar subsample
using the pair-weighted median values of quasar luminosity and redshift.

\subsection{Correcting for Fiber Collisions}\label{sec:fiber_collision}

Due to restrictions of fiber placement during the BOSS survey, two targets
separated by less than 62\arcsec\ (corresponding to $\sim 0.44\,h^{-1}{\rm
Mpc}$ transverse comoving distance at $z=0.55$) cannot be observed
simultaneously on the same plate (tile), but can be both observed on
overlapping plates. The BOSS tiling procedure uses optimized algorithms to
maximize the number of galaxy targets in tile overlap regions, but there are
still $\sim 10\%$ CMASS galaxy targets that do not have a spectroscopic
observation and are lost from the spectroscopic catalog. This fiber collision
effect reduces the number of pairs on small (one-halo) scales and therefore
lowers the clustering strength over these small scales. There are several
schemes to compensate for the preferential loss of quasar-galaxy pairs due to
fiber collisions: upweighting the nearest spectroscopic galaxies that have a
collided target \citep[][]{Anderson_etal_2012}; assigning the photometric
targets a redshift from the nearest spectroscopic neighbor
\citep[e.g.,][]{Zehavi_etal_2005}; or using an algorithm that tracks the
tiling geometry and recovers the true small-scale correlation strength
\citep{Guo_etal_2012}.

Here we decided to use the upweighting scheme to recover the small-scale
cross-correlation signal. In the case of our cross-correlation study, the
spectroscopic observations of BOSS galaxies are completely independent of the
spectroscopic observations of the low-$z$ SDSS-I/II quasars\footnote{This
situation is different from the cross-correlation between galaxies and
quasars from the SDSS-I/II survey, where there is fiber collision between
quasar targets and galaxy targets.}, as the BOSS survey never places a fiber
on a known low-redshift quasar \citep{Ross_etal_2012}. The upweighting scheme
is thus equivalent to the nearest neighbor scheme such that both methods
provide the maximum compensation for pair loss due to fiber collision. The
information on the galaxy weights for fiber-collision (and a smaller fraction
due to redshift failures) corrections is taken from the DR10 CMASS sample.

\subsection{Random Catalogs, Correlation Function Estimators,
  and Error Estimation}

We generate random catalogs for the CMASS galaxy sample with the same angular
geometry and redshift distribution as the data. The spectroscopic
completeness $f_s$ (i.e., fraction of targets with fibers assigned) is a
function of sectors \citep[see e.g.,][for the definition of
sectors]{Blanton_etal_2003}, and is taken into account by upweighting the
galaxy points during pair counting. We already account for fiber collisions,
so the spectroscopic completeness here does not include objects lost to fiber
collisions.

We estimate the 1D and 2D redshift space correlation functions $\xi_s(s)$ and
$\xi_s(r_p,\pi)$ using the simple estimator \citep[][DP]{Davis_Peebles_1983}:
$QG/QR -1$, where $QG$ and $QR$ are the normalized numbers of quasar-galaxy
and quasar-random pairs in each scale bin, $s$ is the pair separation in
redshift space, and $r_p$ ($\pi$) is the transverse (radial) separation in
redshift space.  We shall comment further on this choice below. To reduce the
effects of redshift distortions, we use the projected correlation function
\citep[e.g.,][]{Davis_Peebles_1983}
\begin{equation}
  w_p(r_p) = 2\int_0^\infty d\pi\ \xi_s(r_p,\pi)\ .
\end{equation}
In practice we integrate $\xi_s(r_p,\pi)$ to $\pi_{\rm max}=70\,h^{-1}$Mpc,
where the result is already converged for the scales considered in this paper.
This upper-limit of $\pi_{\rm max}$ will be taken into account in
our subsequent modeling.
For our fiducial $\xi_s(r_p,\pi)$ grid we use a logarithmic binning in $r_p$
with $\Delta\log r_p=0.125$ starting from $r_{p,{\rm min}}=0.1\,h^{-1}$Mpc
and a linear binning in $\pi$ with $\Delta\pi=5\,h^{-1}$Mpc.

There are different methods to estimate the statistical errors of the
correlation function measurement, either internally using bootstrap or
jackknife resampling, or externally using mock catalogs \citep[for a
discussion, see, e.g.,][]{Norberg_etal_2009}. Here we adopt the jackknife
resampling method \citep[as was done in, e.g.,][]{Scranton_etal_2002,
Zehavi_etal_2005,Shen_etal_2007b}: we divide the clustering samples into
$N_{\rm jack}$ spatially contiguous regions with equal area, and create
$N_{\rm jack}$ jackknife samples by excluding each of these regions in turn.
We create our jackknife samples using the pixelization scheme of
STOMP\footnote{\texttt{http://code.google.com/p/astro-stomp/}}, which has
been used in other studies \citep[e.g.,][]{McBride_etal_2011}. We measure the
correlation function for each of these jackknife samples, and the covariance
error matrix is estimated as:
\begin{equation}
{\rm Cov}(i,j)=\frac{N_{\rm jack}-1}{N_{\rm jack}}\sum_{l=1}^{N_{\rm jack}}(\xi_i^l-\bar{\xi}_i)(\xi_j^l-\bar{\xi}_j)\ ,
\end{equation}
where indices $i$ and $j$ run over all bins in the correlation function, and
$\bar{\xi}$ is the mean value of the statistic $\xi$ over the jackknife
samples. The covariance matrix is generally dominated by the diagonal
elements except for the large-scale bins, where correlations between adjacent
$\xi$ bins become important due to common objects in these bins.

We settled on 50 jackknife samples to estimate the covariance matrix. The
normalized covariance matrix (also known as the correlation matrix) is
defined as:
\begin{equation}
  {\rm Cov_{norm}}(i,j)=\frac{{\rm Cov}(i,j)}{\sigma_i\sigma_j}\ ,
\end{equation}
where $\sigma_i^2\equiv {\rm Cov}(i,i)$ is the diagonal element of the
covariance matrix. By default we will use the full covariance matrix in our
model fitting unless otherwise stated. Further discussions on error
estimations and jackknife sampling are presented in the appendix.


\begin{figure}
 \centering
 \includegraphics[width=0.9\linewidth]{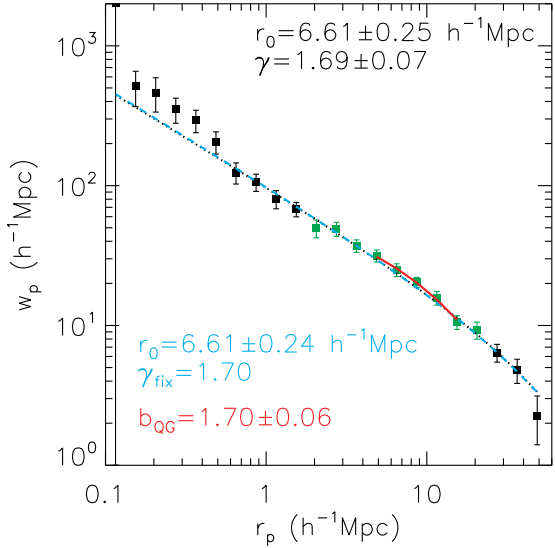}
 \caption{Projected cross-correlation function for the full quasar and CMASS galaxy cross-correlation sample.
 The black and cyan lines are the best-fit power-law model for the scale range $r_p=2-25\,h^{-1}$Mpc with
 flexible power-law index $\gamma$ and fixed index $\gamma_{\rm fix}=1.7$. The red line is the best fit
 linear bias model (i.e., the linear matter correlation function scaled by a constant bias) for the fitting range $r_p=4-16\,h^{-1}$Mpc.
 All fits were performed using the full covariance matrix. }
 \label{fig:wp_all}
\end{figure}

\begin{figure*}
 \centering
 \includegraphics[width=0.9\textwidth]{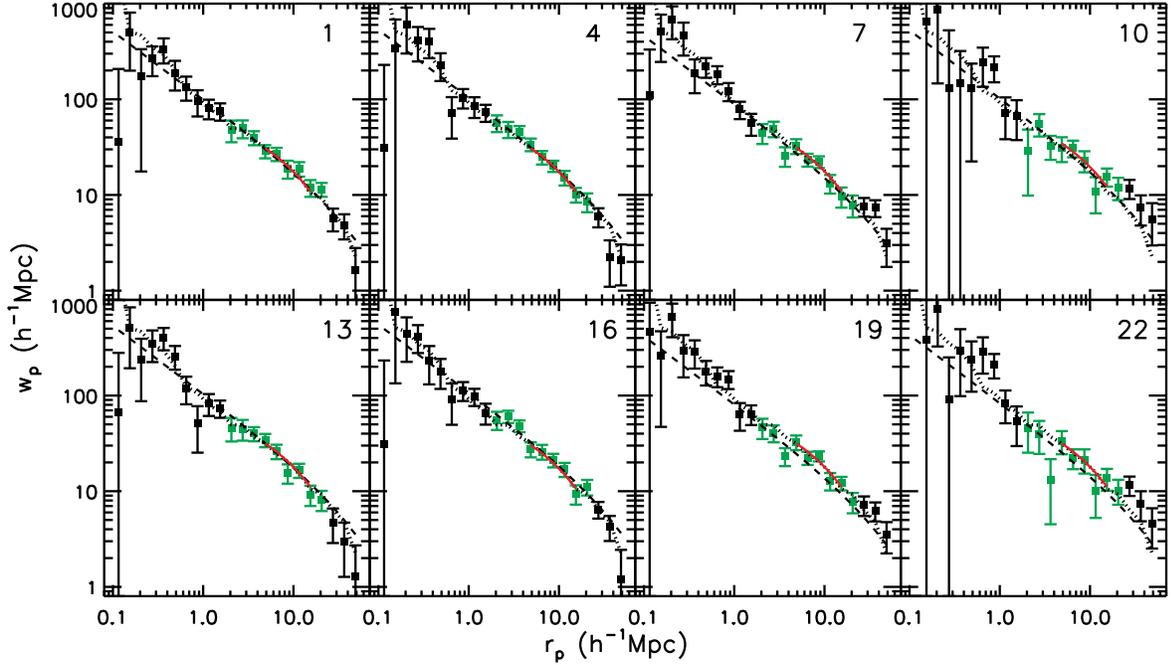}
 \caption{Projected cross-correlation function for the quasar luminosity subsamples with the two
 luminosity divisions (see Fig.\ \ref{fig:mi_z_dist}). The data points are
 measurements for that bin, with green symbols (within $2<r_p<25\,h^{-1}{\rm Mpc}$) indicating those used in the
 power-law model fitting. The $w_p$ data for the full sample is shown in dotted lines as a
 reference. The black dashed lines are the power-law fit to the fitting range
 $r_p=2-25\,h^{-1}$Mpc with fixed slope $\gamma=1.7$, and the red lines are the linear matter
 correlation function scaled by the best-fit linear bias $b_{QG}$ over the fitting range $r_p=4-16\,h^{-1}$Mpc.
 The sample number is marked in each panel (see Table \ref{tab:summary}
 for sample information). }
 \label{fig:wp_lbin}
\end{figure*}

\begin{figure*}
 \centering
 \includegraphics[width=0.9\textwidth]{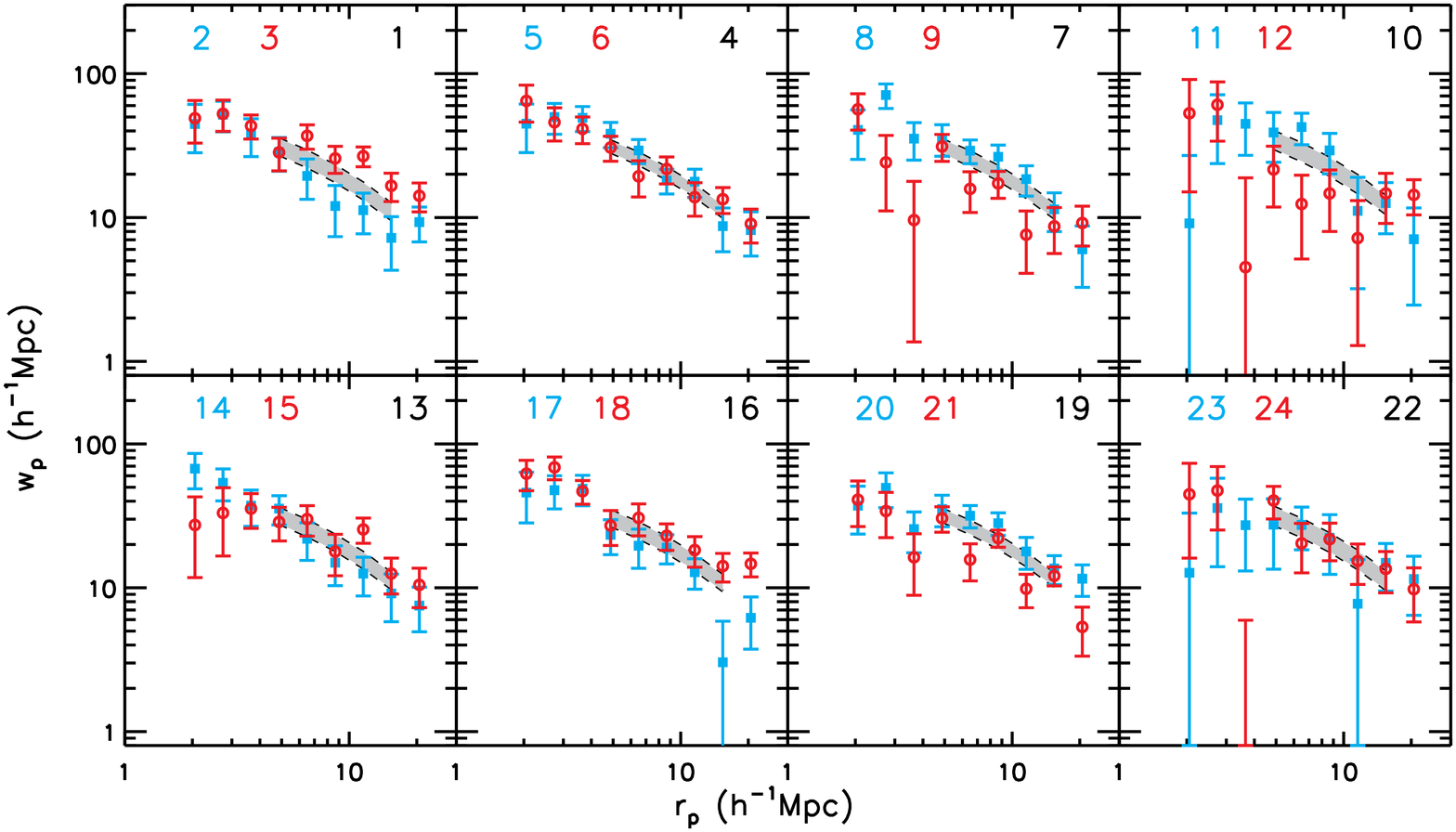}
 \caption{Projected cross-correlation function for the $L-z$ subsamples with the two
 luminosity divisions. In each panel, the black shaded region is the 1$\sigma$ range of the biased linear matter
 correlation function derived from the best-fit to the luminosity subsample indicated by the black number. The cyan
 and red points are the results for the two $L-z$ subsamples of each luminosity subsample. For clarity we only show
 the data points over the $2<r_p<25\,h^{-1}{\rm Mpc}$ range. }
 \label{fig:wp_zbin}
\end{figure*}

\begin{table}
\centering
\caption{Measurements of the cross-correlation function $w_p$ for
the full sample and subsamples. The second column lists the total raw number
of $QG$ pairs in a given $r_p$ bin with $\pi\le 70\,h^{-1}{\rm Mpc}$, which
can be used as a rough estimate of the robustness of the sample statistics.
The last column lists the diagonal errors of the $w_p$ measurements, and the
normalized covariance matrices are provided in Table \ref{tab:CM}. A portion
is shown here for its content. The table is available in its entirety in the
electronic version of this paper. } \scalebox{1.0}{
\begin{tabular}{lcccc}
\hline
sample & $r_p$ & $QG$ & $w_p$ & $\sigma_{w_p,{\rm diag}}$ \\
\# & ($h^{-1}$Mpc) & & ($h^{-1}$Mpc) & ($h^{-1}$Mpc) \\
\hline
0 & $0.1155$ & 12 & $2061.9628$ & $2440.1567$\\
  & $0.1540$ & 23 & $513.5358$ & $145.3023$\\
  & $0.2054$ & 38 & $464.1206$ & $127.6868$\\
\hline
\end{tabular}}\label{table:wp}
\end{table}

\begin{table}
\centering \caption{Quasar linear bias derived from $b_{QG}$ and $b_G$. The
error bars are simply propagated from $b_{QG}$ and $b_G$ neglecting
covariance. We only tabulated the results for the luminosity subsamples
(e.g., the results for the $L-z$ subsamples are too noisy to be useful). Note
that the data for the most luminous subsample (s4) are a subset of the less
luminous subsample (s3), so the bias measurements in these two bins are not
independent. } \scalebox{1.0}{
\begin{tabular}{lccc}
\hline
sample & $\langle z\rangle$ & $\langle M_i\rangle$ & $b_Q$ \\
\hline
Full          & 0.532 & $-24.055$  & $1.38\pm0.10$   \\
div1\_s1\_z1  & 0.533 & $-23.675$  & $1.35\pm0.18$   \\
div1\_s2\_z1  & 0.531 & $-24.000$  & $1.36\pm0.13$   \\
div1\_s3\_z1  & 0.533 & $-24.727$  & $1.37\pm0.15$   \\
div1\_s4\_z1  & 0.533 & $-25.406$  & $1.52\pm0.21$   \\
div2\_s1\_z1  & 0.484 & $-23.564$  & $1.36\pm0.17$   \\
div2\_s2\_z1  & 0.547 & $-24.040$  & $1.35\pm0.17$   \\
div2\_s3\_z1  & 0.578 & $-24.741$  & $1.42\pm0.15$   \\
div2\_s4\_z1  & 0.579 & $-25.417$  & $1.42\pm0.20$   \\
\hline
\end{tabular}}\label{table:bQ}
\end{table}

\section{The Cross Correlation Function}\label{sec:cf}

\subsection{The whole quasar sample}\label{sec:cf:whole}

We show the projected correlation function $w_p$ for the full quasar and
CMASS galaxy samples in Fig.\ \ref{fig:wp_all}, and tabulate the measurements
in Table \ref{table:wp}.
Much of our focus will be on the larger scales measurements, but it can be
seen that we have a good detection of clustering to quite small scales.
In particular, there are 842 $QG$ pairs within $r_p<1\,h^{-1}{\rm Mpc}$ and
$\pi<70\,h^{-1}{\rm Mpc}$, allowing a fair estimate of the small-scale
(one-halo) cross-correlation.

We fit the measured CCF with a power-law model $\xi(r)=(r/r_0)^{-\gamma}$
over the projected scales $2<r_p<25\ h^{-1}{\rm Mpc}$ to quantify the
clustering strength on intermediate-scales. We can also estimate a linear
bias $b_{QG}$, i.e.,
\begin{equation}
w_p = w_{p,{\rm matter}}b_{QG}^2\ ,
\end{equation}
where $w_{p,{\rm matter}}$ is the correlation function of the underlying
matter at the redshift of interest, and $b_{QG}^2\approx b_{Q}b_G$ where
$b_Q$ and $b_G$ are the linear biases for the quasar and CMASS samples
respectively.


To estimate the linear bias $b_{QG}$, we use the linear matter correlation
function computed using the linear power spectrum in
\citet{Eisenstein_Hu_1999} under the adopted cosmology, estimated at the
pair-weighted median redshift of the cross-correlation samples. Our
investigations using mock catalogs (see \S\ref{sec:disc3}) show that on
scales $r_p\lesssim 4\,h^{-1}$Mpc nonlinear and one-halo effects start to
affect the linear bias, while at $r_p\gtrsim 15\,h^{-1}$Mpc residual redshift
space distortion (RSD) effects start to become important. Thus we narrow the
fitting range to $r_p=[4,16]\,h^{-1}$Mpc to estimate the linear bias, where a
scale-independent linear bias seems to be a good approximation (within
$10\%$). Although we lose statistical power by excluding data points (i.e.,
only 5 bins of scale are used in the fitting), this procedure is preferred to
avoid scales where non-linear effects, scale-dependent bias, and RSDs may
affect the linear bias estimate. Nevertheless we tested varying $r_p$
boundaries within $[1,50]\,h^{-1}{\rm Mpc}$ in the fitting and found all
derived $b_{QG}$ values are consistent within 1$\sigma$, thus our estimate of
$b_{QG}$ is robust against this detail.

The correlation function is well fitted by a power-law model with
$r_0=6.61\pm 0.25$ and $\gamma=1.69\pm0.07$ over the scales of
$2<r_p<25\,h^{-1}{\rm Mpc}$ ($\chi^2/{\rm dof}=6.54/7$). On smaller scales,
the correlation function significantly deviates from the best-fit power-law
model derived from larger scales, and requires explicit modeling of the
one-halo term. The fact that we detect significant clustering at $r_p\lesssim
1\,h^{-1}{\rm Mpc}$ indicates that there are a population of satellite hosted
quasars and CMASS galaxies in the cross-correlation sample (see discussions
in \S\ref{sec:disc}).

The linear bias for the full cross-correlation sample from our simple fitting
is $b_{QG}=1.70\pm 0.06$. In order to derive the quasar linear bias $b_{Q}$
we need to know the linear bias of CMASS galaxies $b_{G}$. For this purpose
we have measured the auto correlation function (ACF) for the CMASS galaxy
sample using the standard DP estimator, and used the same fitting procedure
to estimate $b_{G}$. However, we found that the best-fit $b_G$ value does
depend on the exact fitting range, given the substantially smaller
statistical errors from the ACF measurement. To reduce the risk of
contamination from small-scale non-linear clustering and large-scale
redshift-space distortion, we fit the CMASS ACF over the same scale range
($r_p=[4,16]\,h^{-1}{\rm Mpc}$) as for the CCF data, and derive $b_G=2.10\pm
0.02$. Within this fitting range, the ratio of the CCF to the galaxy ACF is
roughly constant, allowing use of the relation $b_{QG}^2=b_Qb_G$ to derive
the quasar linear bias. The inferred quasar linear bias is $b_{Q}\sim 1.38\pm
0.10$, consistent with the estimated $b_{Q}\sim 1.3\pm0.2$ from the SDSS
quasar auto-correlation function measured at $\langle z\rangle\sim 0.5$
\citep[e.g.,][]{Shen_etal_2009a}. This linear bias is also consistent with
the value derived using the HOD approach described in \S\ref{sec:disc2} and
with the bias of the mock catalogs (which show a slight, slow decrease of the
inferred bias from $4\,h^{-1}$Mpc to $16\,h^{-1}$Mpc).

Our derived CMASS galaxy bias value is somewhat larger than the estimated
value of $1.8-2$ in other ACF studies of CMASS galaxies
\citep[e.g.,][]{White_etal_2011,Nuza_etal_2012}, but is consistent with that
derived in \citet{Guo_etal_2012b} based on the DR9 CMASS sample. This result
is at least partly caused by the different methodology in estimating the
bias. We also compared our ACF measurement directly with those reported in
other studies
\citep[e.g.,][]{White_etal_2011,Anderson_etal_2012,Nuza_etal_2012}; our
measurement is systematically higher by $\sim 10\%$ over
$r_p=4-16\,h^{-1}{\rm Mpc}$ scales. To resolve this discrepancy we performed
extensive tests upon our galaxy sample and the samples used in other studies,
and found that this systematic difference is largely due to the usage of
additional galaxy weights in the other studies. While there are good reasons
to use those weights in these studies, it is not clear that they are
applicable to our cross-correlation measurements. On the other hand, we
tested the difference of using the simple DP estimator and the more robust
Landy-Szalay \citep[][LS]{Landy_Szalay_1993} estimator, and found that the DP
estimator over-estimates $w_p$ by only $<2\%$ below $r_p=10\,h^{-1}{\rm Mpc}$
and by $\sim 10\%$ at $r_p\sim 40\,h^{-1}{\rm Mpc}$, which means the
difference caused by using the simple DP estimator is negligible. In general
the statistical errors tabulated in Table \ref{tab:summary} are significantly
smaller than the systematic uncertainties in the galaxy bias estimation.
Nevertheless, regarding the detection of the luminosity dependence of quasar
bias, the exact value of the galaxy bias is not critical.




\begin{figure*}
 \centering
 \includegraphics[width=0.45\textwidth]{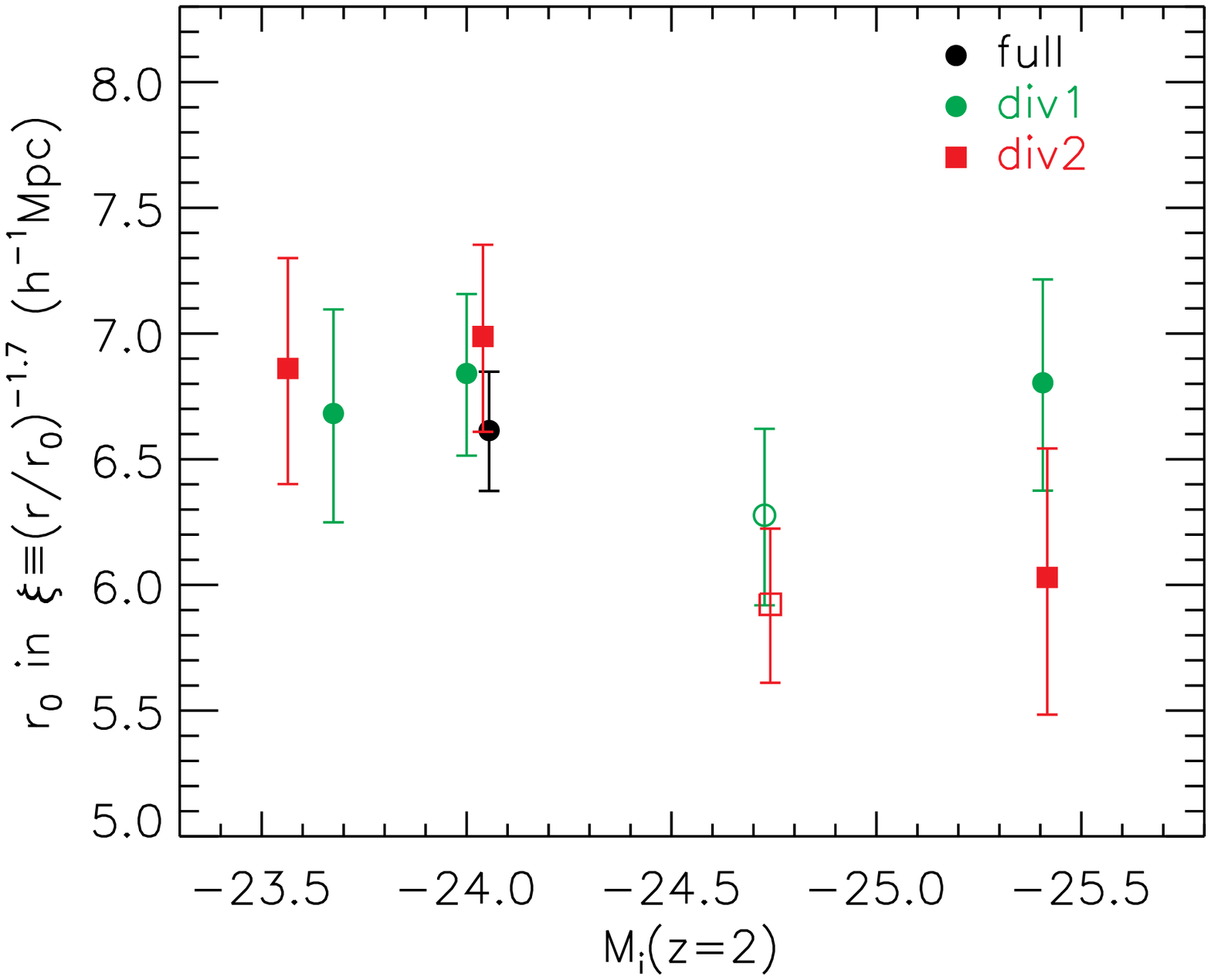}
 \includegraphics[width=0.45\textwidth]{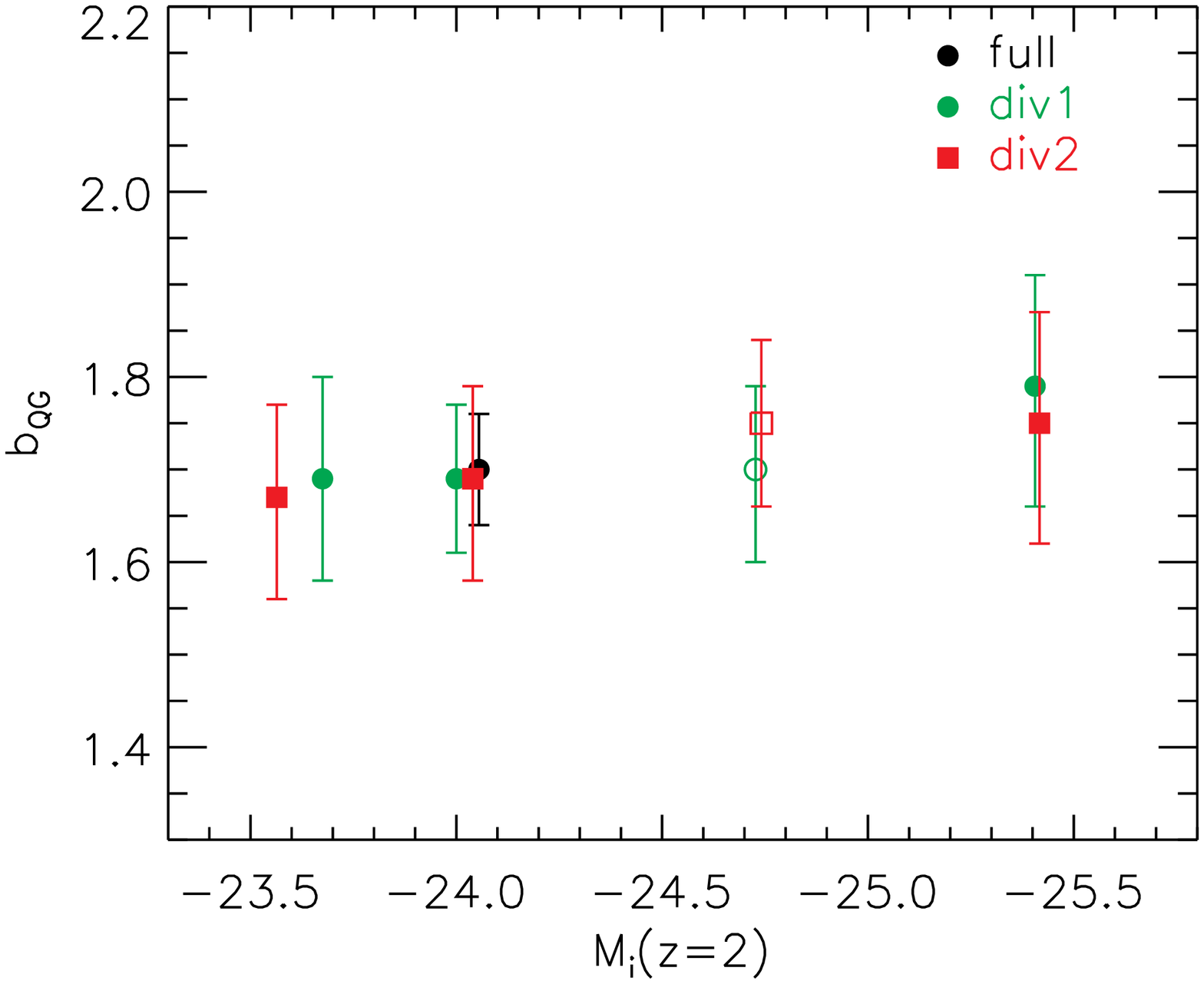}
 \caption{The strength of the cross-correlation in terms of $r_0$ (left) from the power-law model fits and linear bias
 $b_{QG}$ (right) for different luminosity subsamples. These estimates are tabulated in Table \ref{tab:summary}. We use open symbols
 for the second most luminous subsample (s3) in the two divisions to indicate the fact that it contains the most luminous subsample (s4). }
 \label{fig:r0_b0_plot}
\end{figure*}

\begin{figure*}
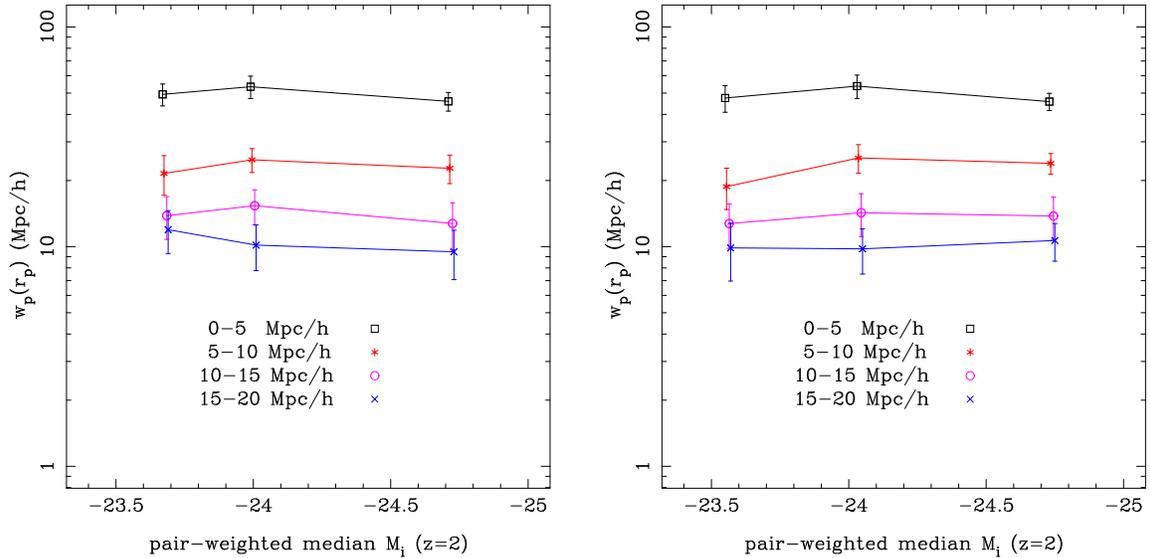

 \centering
 \includegraphics[width=0.4\linewidth]{wp_avg_div1.eps}
 $\qquad$
 \includegraphics[width=0.4\linewidth]{wp_avg_div2.eps}
    \caption{Clustering in larger (averaged) bins as a function of their median
      pair-weighted magnitude, for Division 1 (left) and Division 2 (right).
      Only the first three luminosity subsamples in each division are shown.
      The errors denote the 1$\sigma$ uncertainty from jackknife re-sampling with 50 regions.
      This demonstrates that the shape and amplitude of the cross-correlation function
      show no significant variation for different quasar luminosity subsamples.
    }
    \label{fig:wp_avg}
\end{figure*}

\begin{figure*}
 \centering
 \includegraphics[width=0.45\textwidth]{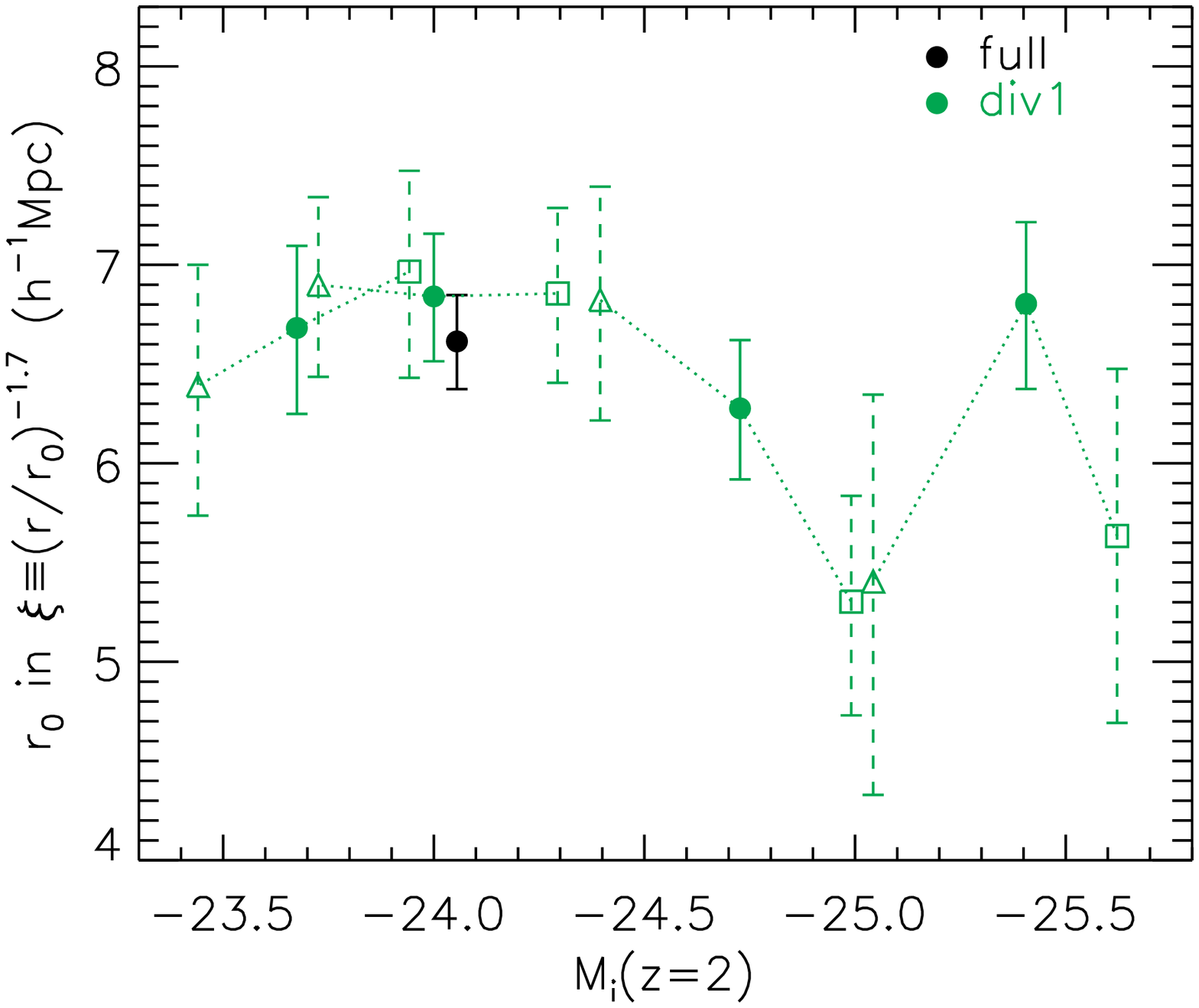}
 \includegraphics[width=0.45\textwidth]{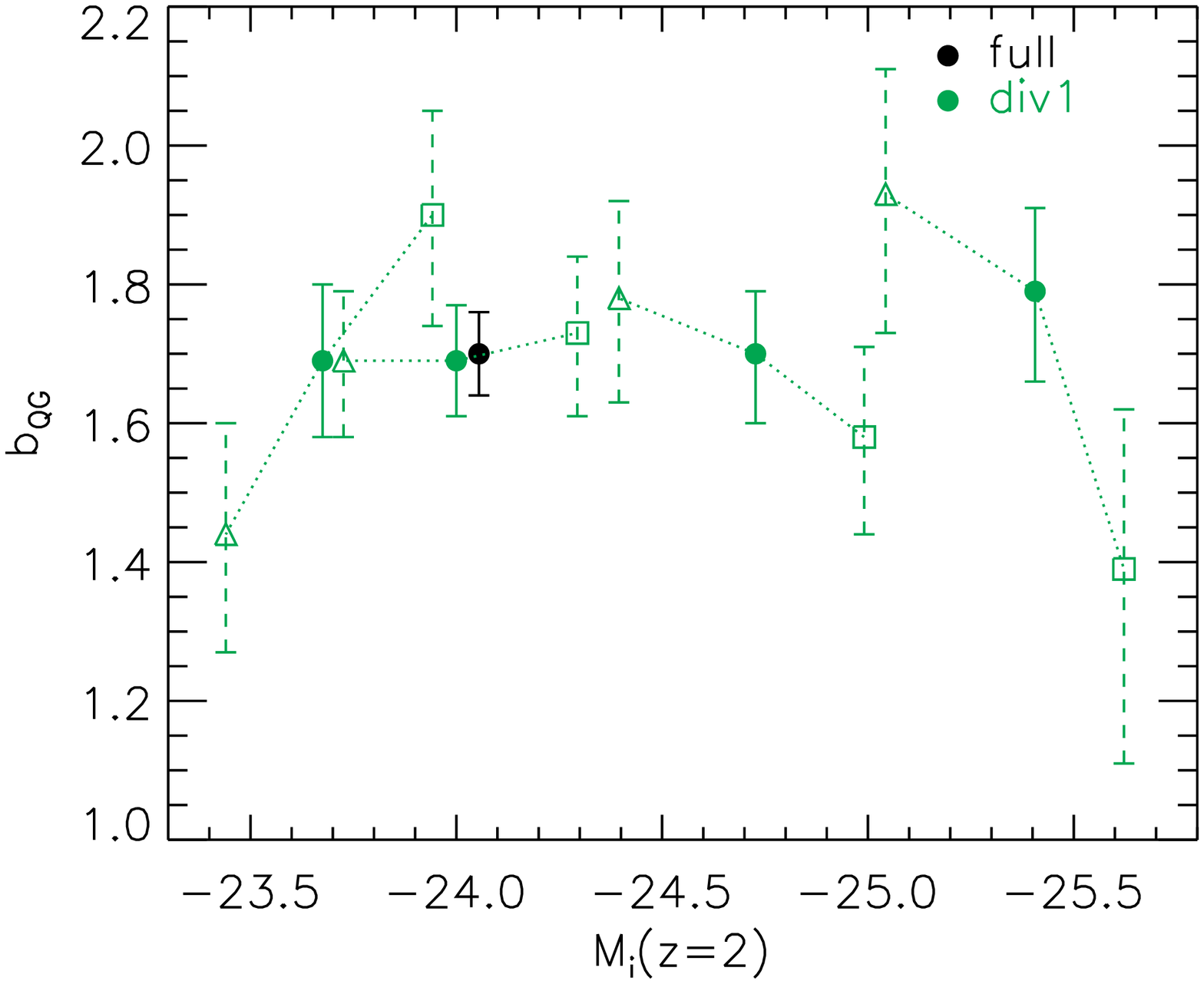}
 \includegraphics[width=0.45\textwidth]{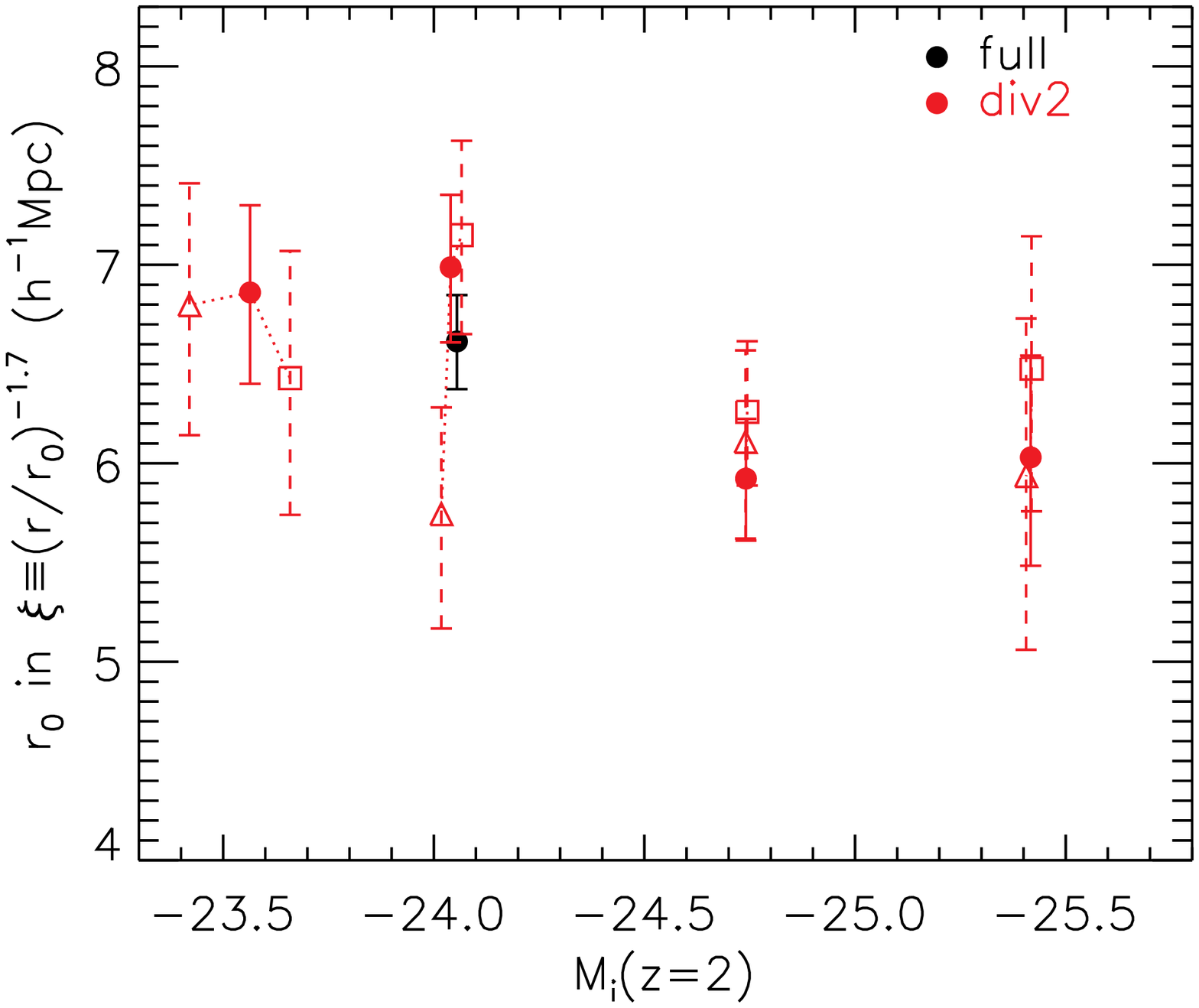}
 \includegraphics[width=0.45\textwidth]{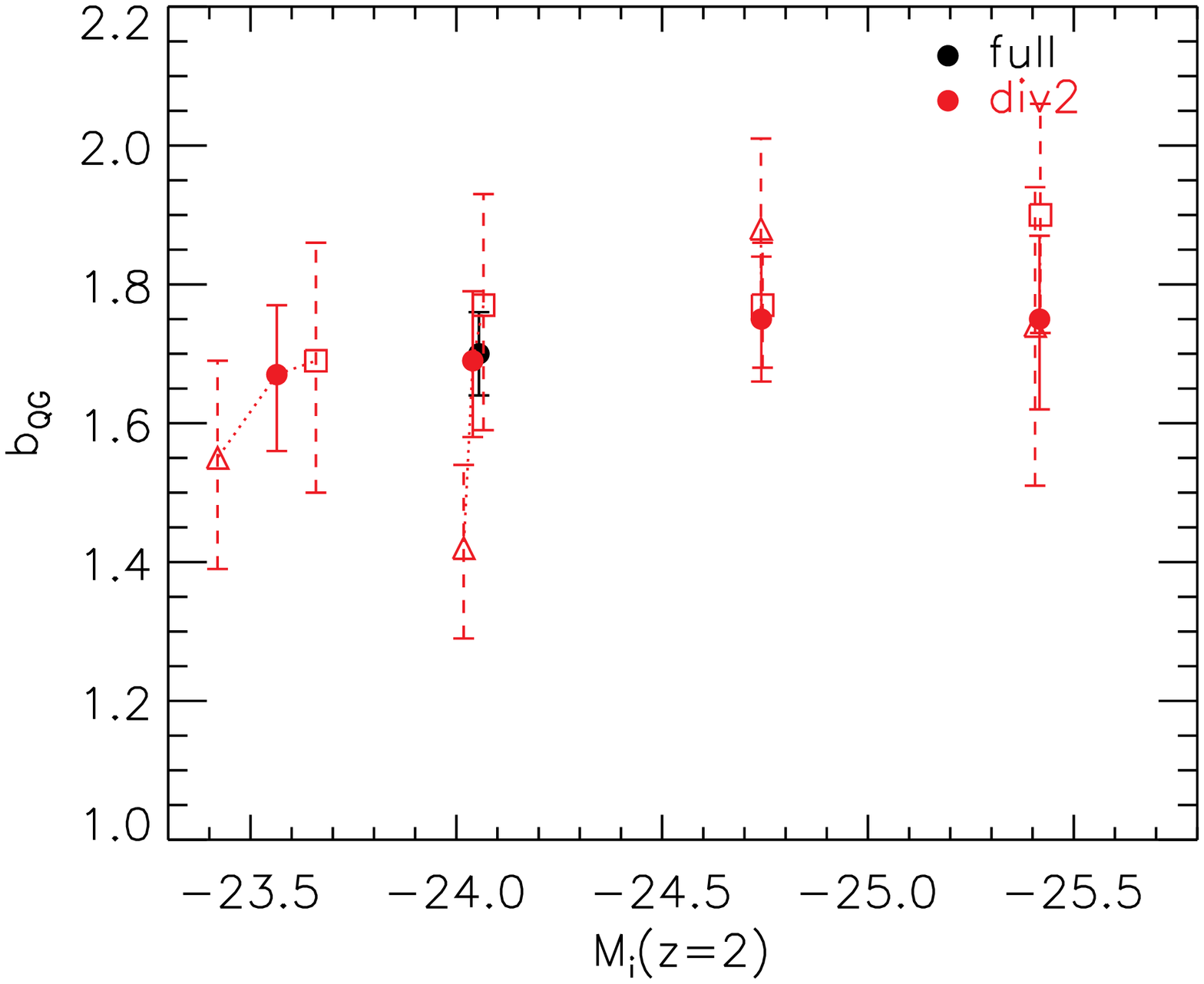}
 \caption{The strength of the cross-correlation in terms of $r_0$ and linear bias $b_{QG}$. For each luminosity
 subsample we further plot the results of the two redshift subsamples, connected by the dotted lines. No
 redshift difference is detected given the large error bars. }
 \label{fig:r0_b0_zplot}
\end{figure*}

\begin{figure*}
 \centering
 \includegraphics[width=0.48\textwidth]{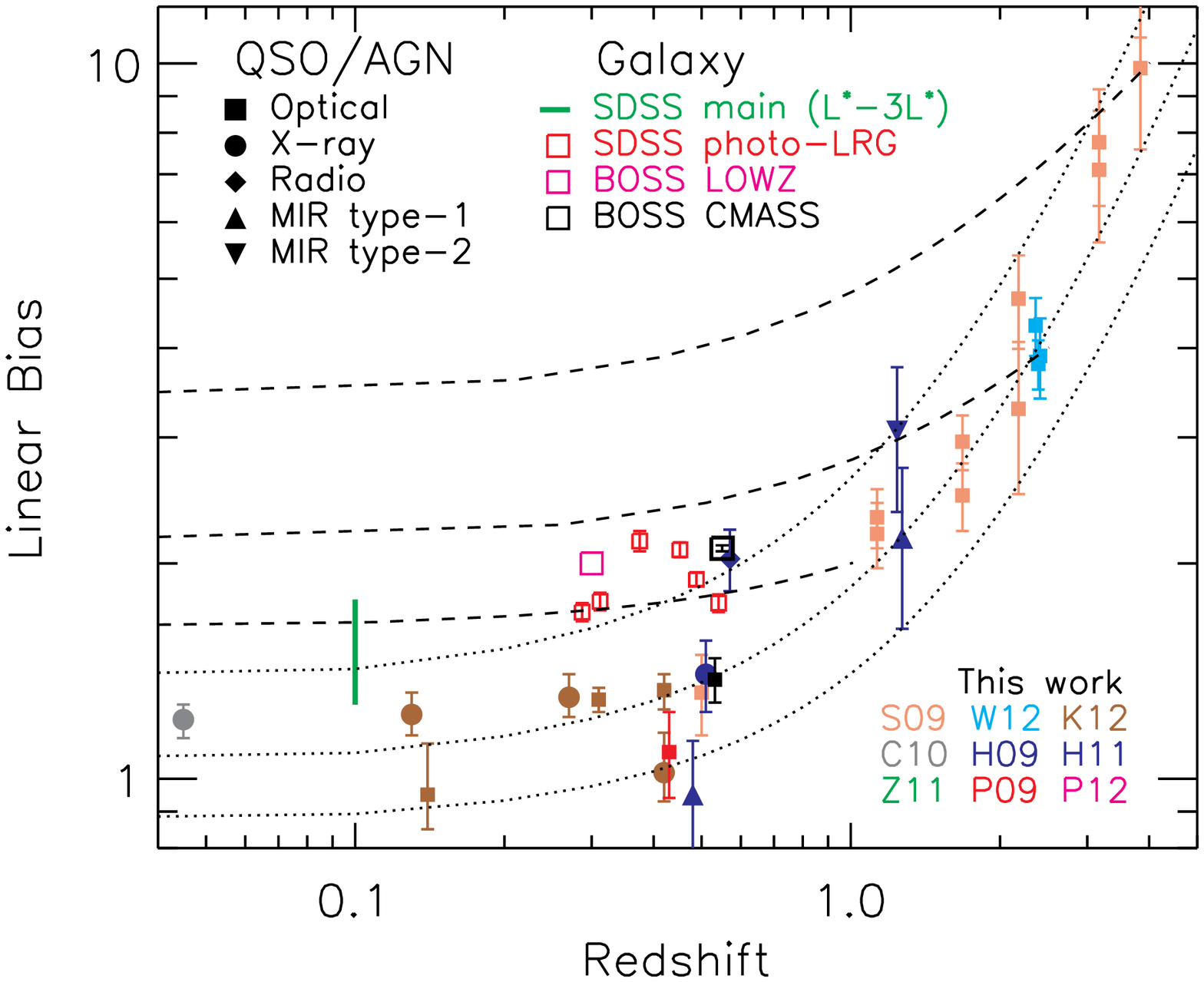}
 \includegraphics[width=0.48\textwidth]{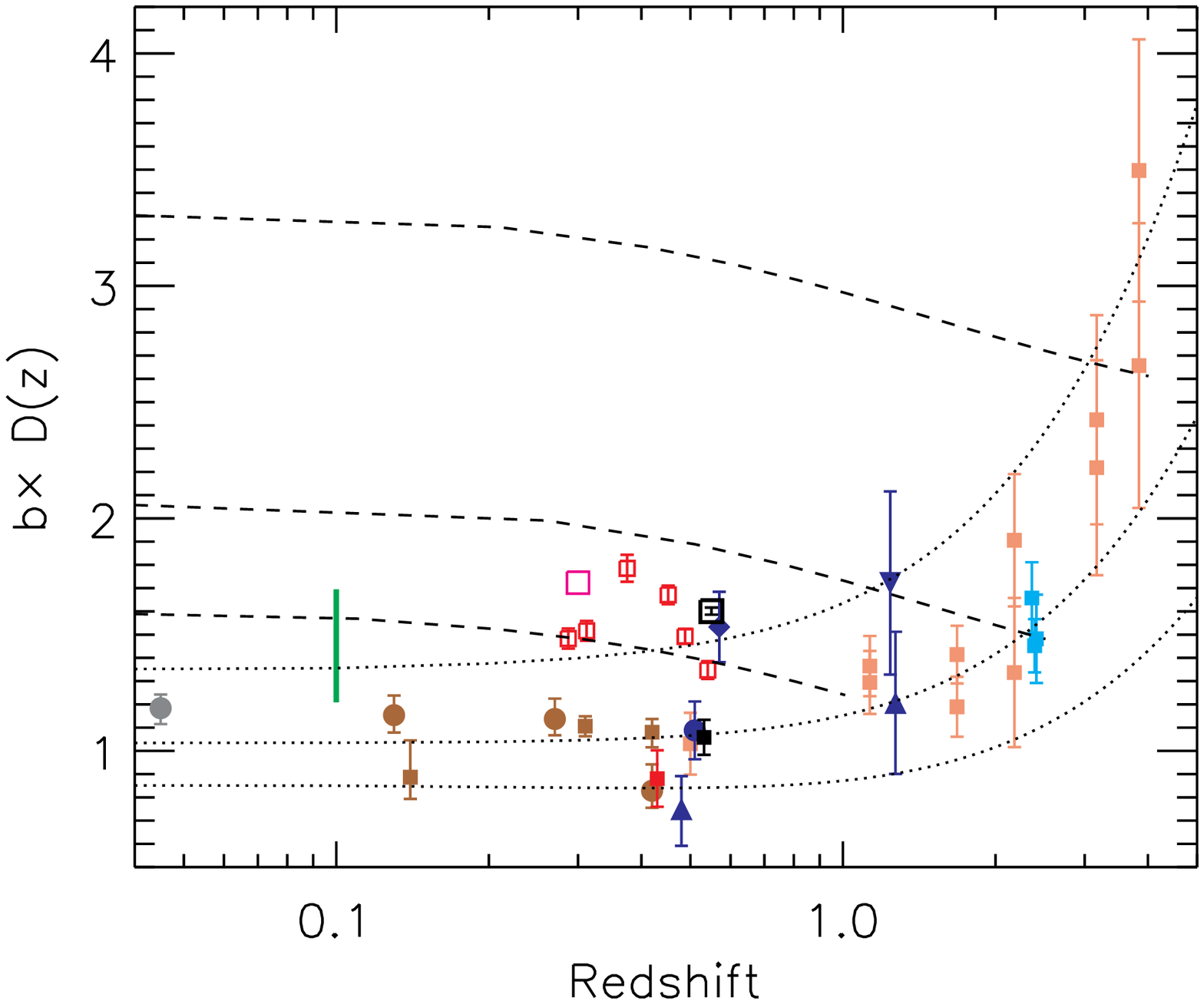}
 \caption{{\em Left:} comparison of the linear bias derived for different tracer samples. The
 solid symbols are for quasars and AGNs, while the open symbols and the green vertical line segment are for
 galaxies. Measurements are from \citet[][S09]{Shen_etal_2009a}, \citet[][W12]{White_etal_2012},
 \citet[][K12]{Krumpe_etal_2012}, \citet[][C10]{Cappelluti_etal_2010},
 \citet[][H09]{Hickox_etal_2009}, \citet[][H11]{Hickox_etal_2011}, \citet[][Z11]{Zehavi_etal_2011},
 \citet[][P09]{Padmanabhan_etal_2009}, and \citet[][P12]{Parejko_etal_2012}. The three
dotted lines are the halo linear bias estimated using the recipes provided in
\citet{Tinker_etal_2005} for
 halo masses $M_h=1,4,16\times 10^{12}\,h^{-1}M_\odot$. Note that different fitting formula for the halo bias will yield
  slightly different results \protect\citep[e.g.,][]{Sheth_Mo_Tormen_2001}. The
  three dashed lines are the predicted bias evolution for a passive
  population \citep[e.g.,][]{Fry_1996,Mo_White_1996,Hopkins_etal_2008a}, started at three arbitrary high redshifts and matched
  to the measured linear bias of quasars at these redshifts. These biases derived in different work used different methods, and while
  they often agree within the reported error bars, there are cases when the
  reported error bars underestimate the systematic uncertainty in determining
  the bias \citep[e.g.,][]{Padmanabhan_etal_2009,Krumpe_etal_2012},
  especially when the statistical uncertainty is small. With these caveats in
  mind, this figure suggests that quasars at different redshifts reside in halos with typical masses of a few $10^{12}\,h^{-1}M_\odot$, and as such
  low-redshift quasars are not the descendants of their high-redshift
  counterparts, which would have evolved into more massive systems. The
  massive galaxies at $z\lesssim 0.5$ in the SDSS samples typically reside in $\sim 10^{13}\,h^{-1}M_\odot$
  halos, and could be the descendants of $z\sim 1$ quasars. {\em Right}: Same as the left panel, but with the product of the linear
  bias and the linear growth factor $D(z)$ as the $y$-axis. Thus constant large-scale clustering is denoted by horizontal lines in this plot. }
 \label{fig:b0_zevo}
\end{figure*}

\begin{figure}
 \centering
 \includegraphics[width=0.9\linewidth]{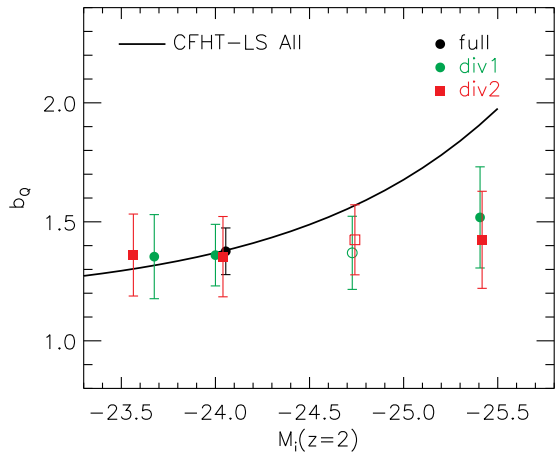}
 \caption{Comparison of the luminosity dependence of quasar bias derived in this work (symbols) with that of
 galaxies in the CFHT-LS sample (black solid line) at $0.4<z<0.6$ \citep[][]{Coupon_etal_2012}. We use open symbols
 for the second most luminous subsample (s3) in the two divisions to indicate the fact that it contains the most luminous subsample (s4).
 To map between quasar luminosity and
 galaxy luminosity we have assumed that the typical quasar luminosity in our sample ($M_i(z=2)=-24.055$) corresponds to the
galaxy luminosity with the same bias. Incidently we get a corresponding galaxy luminosity of $\approx L^*$. Note that the
galaxy biases were derived for luminosity-threshold samples, and we have limited the galaxy luminosity within the range of
$0.15-3L^*$, approximately the range probed by the CFHT-LS sample. The luminosity dependence of quasar bias is apparently
weaker than that of the galaxy bias.}
 \label{fig:b0_qso_gal_comp}
\end{figure}

\begin{figure*}
 \centering
 \includegraphics[height=0.4\textwidth]{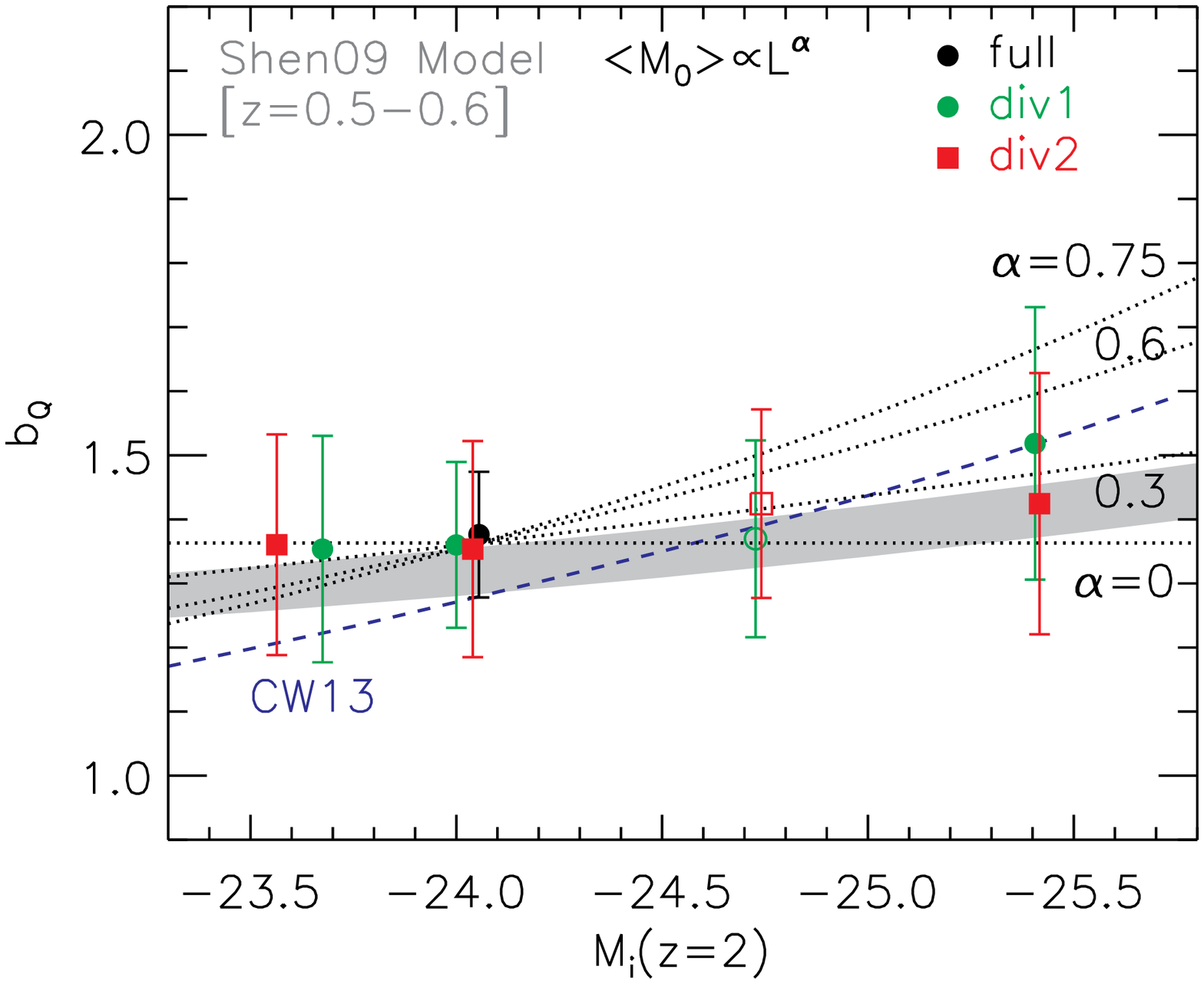}
 \includegraphics[height=0.4\textwidth]{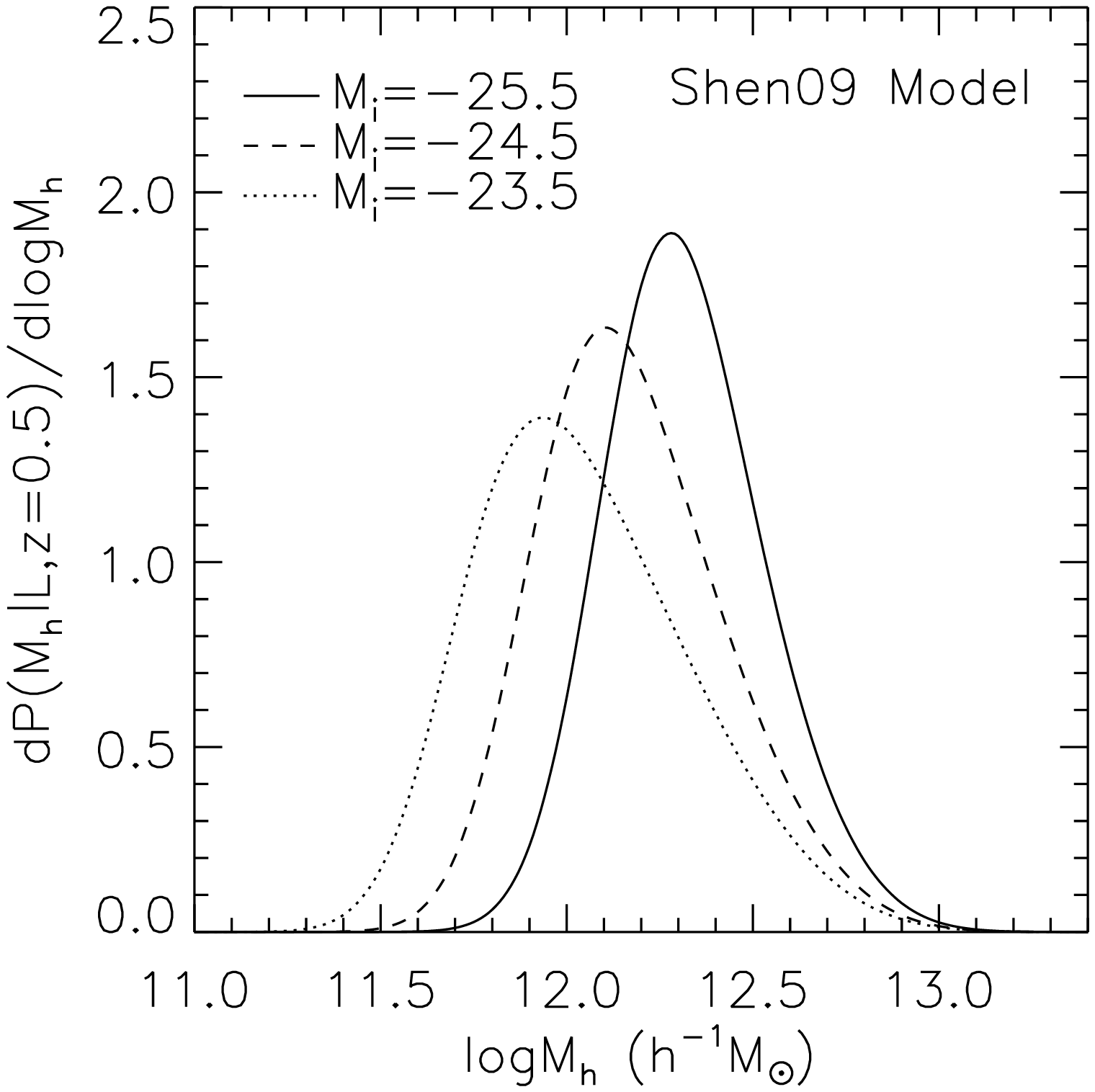}
 \caption{{\em Left:} Comparisons between several model predictions and our measurement of the
 luminosity dependence of quasar large-scale linear bias. We use open symbols
 for the second most luminous subsample (s3) in the two divisions to indicate the fact that it contains the most luminous subsample (s4).
 For the dotted lines (i.e., power-law
 models with $\alpha=0,0.3,0.6,0.75$), the predictions are generated using the \citet{Tinker_etal_2005} halo bias
 formula at $z=0.53$, and normalized such that they are close to the measured bias for the
 full quasar sample. The gray band is the prediction at $z=0.5-0.6$ from the \citet{Shen_2009} model, and the
 blue dashed line is the prediction at $z=0.55$ from the fiducial model in \citet[][CW13]{CW13} neglecting the satellite contribution (which serves
 to increase the bias in the fainter bins by about 5\% while leaving the bright bins almost unchanged).
 {\em Right:} The distribution of host halo mass at fixed quasar
 luminosity from the Shen (2009) model, estimated at $z=0.5$. }
 \label{fig:b0_model_comp}
\end{figure*}

\subsection{Quasar subsamples divided in luminosity}\label{sec:cf:lum}

Fig.\ \ref{fig:wp_lbin} shows the resulting cross-correlation function for
each quasar luminosity subsample (i.e., no dividing in redshift), and
comparison with that for the full sample. For each luminosity subsample we
show in Fig.\ \ref{fig:wp_zbin} the results for the $L-z$ subsamples. All the
measurements are tabulated in Table 2.

Our current samples do not have a sufficient number of small-scale $QG$ pairs
($r_p\lesssim 1\, h^{-1}{\rm Mpc}$) to probe the clustering difference on
these one-halo scales when dividing our quasar sample in luminosity. To
quantify the luminosity-dependence of the large-scale clustering strength we
fit $w_p$ in the range of $2<r_p<25\,h^{-1}{\rm Mpc}$ with the power-law
model and in the range of $4<r_p<16\,h^{-1}$Mpc with the linear-bias model.
For the power-law model we fix the slope to be $\gamma=1.7$, consistent with
the best-fit slope for the full cross-correlation sample. The amplitude of
the clustering is therefore measured by the best-fit correlation length $r_0$
and linear bias $b_{QG}$.

The best-fit values of $r_0$ and linear bias $b_{QG}$ for different CCF
subsamples are shown in Fig.\ \ref{fig:r0_b0_plot} for the four quasar
luminosity subsamples in each division. No significant difference is detected
among these subsamples. In Fig.\ \ref{fig:wp_avg} we present the $w_p$ values
computed over wide linear $r_p$ bins with $\Delta r_p=5\, h^{-1}{\rm Mpc}$
for the four luminosity subsamples in the two divisions. These $w_p$ values
represent the averaged correlation over these wide bins. Again, we see that
while the value of $w_p$ depends on scale, there is no significant difference
in clustering strength between any of the samples on these scales.  Our
sample statistics are insufficient to probe potential luminosity dependence
on $r_p\lesssim 1\,h^{-1}{\rm Mpc}$ scales.

One concern is that for Div 2 the effective redshift is slightly different
for each luminosity subsample, and possible redshift evolution may complicate
the interpretation. However, {the difference in the linear growth factor over
the probed redshift range ($z\sim 0.45-0.65$) is only $\sim 10\%$, and the
evolution in the linear bias $b_G$ of the CMASS galaxy sample over this
redshift range is negligible (see Table \ref{tab:summary}). Thus the effect
of redshift evolution is negligible for our samples, as expected, and we do
not observe a significant difference when we further divide our luminosity
subsamples in redshift (e.g., Fig.\ \ref{fig:r0_b0_zplot}). }

\section{discussion}\label{sec:disc}

The improved measurement of quasar large-scale clustering at $z\sim 0.5$, and
the inferred luminosity dependence of quasar bias, can be used to study the
evolution of the global quasar population and to test cosmological quasar
models while the small-scale cross-correlation probes the immediate
neighborhood of quasars and may hint at the triggering mechanism of quasars.
Since the statistics on the small-scale cross-correlation in the present
study are still not sufficient for detailed studies (see Fig.\
\ref{fig:wp_lbin}), much of our following discussion will focus on the
large-scale quasar bias and its luminosity dependence, although we do attempt
to model the small-scale clustering for the full cross-correlation sample.

Quasars reside in dark matter halos, and the redshift evolution of quasar
bias can be used to understand the cosmic evolution of this population. A
long-lived quasar population may passively evolve into their lower redshift
counterparts with a predicted bias evolution
\citep[e.g.,][]{Fry_1996,Tegmark_Peebles_1998,Mo_White_1996,White_etal_2007,
Hopkins_etal_2007b},
and can be confronted with the observed quasar bias evolution
(see \S\ref{sec:disc1}).

The observed luminosity dependence of quasar bias constrains how well quasar
luminosity correlates with halo mass. In a physical galaxy formation scheme,
there are various correlations among halo, galaxy and BH properties such that
a chain of $L_{\rm qso}\leftrightarrow M_{\rm BH}\leftrightarrow M_{\rm
gal}\leftrightarrow M_{\rm h}$ may form. If the BH mass is more directly
connected to halo mass than to galaxy mass, we expect a simpler version,
$L_{\rm qso}\leftrightarrow M_{\rm BH}\leftrightarrow M_{\rm h}$. In the
simplest scenario, i.e., all quasars are shining at a constant Eddington
ratio, and BH mass linearly correlates with halo mass with no scatter, we
expect a strong luminosity dependence of quasar bias as a result of more
luminous quasars living in more massive halos. In practice, there are
inevitably curvature and scatter among these correlations, which will modify
the resulting luminosity dependence of quasar bias. For instance, quasar
luminosity at fixed BH mass may have a substantial dispersion, as a natural
result from different fueling conditions; BH mass may not perfectly (and
linearly) correlate with halo mass due to diversities in galaxy formation
details. These scatters will produce a distribution of host halo mass at
fixed quasar luminosity; the more these halo masses overlap in different
quasar luminosity bins, the less prominent will be the observed luminosity
dependence of quasar bias. This effect will be further illustrated in the
following discussion.

\subsection{Implications from large-scale clustering}\label{sec:disc1}

Fig.\ \ref{fig:b0_zevo} presents the quasar/AGN bias measured in different
studies and comparisons to the bias of different galaxy samples. The three
dotted lines show the bias of halos with constant halo mass
$M_h=1,4,16\times10^{12}\,h^{-1}{\rm Mpc}$ using the \citet{Tinker_etal_2005}
halo bias formula\footnote{Using alternative halo bias formula calibrated
against simulations will yield slightly different results that are consistent
within a factor of two
\citep[e.g.,][]{Sheth_Mo_Tormen_2001,Cohn_White_2008}.}. The three dashed
lines show the evolution of bias for a passive population of tracers
\citep[e.g.,][]{Fry_1996,Tegmark_Peebles_1998,Mo_White_1996,White_etal_2007,
Hopkins_etal_2007b}.


These different samples probe different redshifts and luminosities, and are
selected with different methods, thus a detailed comparison would be
difficult. Furthermore, these studies used different methodologies to
estimate the linear bias. Although in most cases the bias values derived with
different methods agree to within 1$\sigma$, there are cases where they could
differ significantly \citep[e.g.,][]{Padmanabhan_etal_2009,Krumpe_etal_2012}.
Keeping these caveats in mind, some general conclusions can be drawn from
this figure:
\begin{itemize}
\item Optically selected quasars appear to have a typical halo mass
    between $10^{12}-10^{13}\, h^{-1}M_\odot$
    \citep[e.g.,][]{Croom_etal_2005,Hopkins_etal_2007b,Shen_etal_2009a,
     Shanks_etal_2011} over a wide redshift range. This result implies
     that most low-$z$ quasars are not the descendants of their high-$z$
    counterparts, which would have evolved into systems with relatively
    higher bias at low redshift.


\item There is no significant difference in the clustering strength
    between optical quasar samples and several X-ray selected AGN samples
    at the same redshift \citep[e.g.,][]{Krumpe_etal_2012}. However, we
    note that these X-ray AGN samples only probe slightly fainter
    luminosities than the optical quasar samples, thus both types of
    active SMBHs are likely drawn from a similar population, and
    therefore should trace a comparable halo mass range. There may be some
    hints that radio-selected AGNs have higher clustering than optical
    quasars and X-ray selected AGNs
    \citep[e.g.,][]{Wake_etal_2008,Hickox_etal_2009,Donoso_etal_2010}.

\item The galaxy populations from SDSS and BOSS are significantly more
    clustered than quasars/AGNs at the same redshift. By selection these
    galaxy samples are at the massive end of the galaxy population. Thus
    most low-$z$ quasars are not shining within these massive galaxies.
    These massive galaxies may have experienced a brief quasar phase in
    the past to build up the central SMBH mass, and are therefore likely
    the descendants of high-$z$ quasar host galaxies.


\end{itemize}

{At $z\sim 0.5$, the average stellar mass of the CMASS galaxy sample is $\sim
2\times10^{11}\,M_\odot$ (Maraston et~al.\ 2012). This value corresponds to a
black hole mass of $\sim 4\times 10^8\,M_\odot$ using the local $M_{\rm
BH}-M_{\rm bulge}$ relation in \citet{Marconi_Hunt_2003} and assuming all the
stellar mass is in the bulge for CMASS galaxies. The average BH mass of the
SDSS quasars is estimated to be $\sim 4\times10^{7}\,M_\odot$ (assuming unity
Eddington ratio) or $\sim 3\times10^8\,M_\odot$ (virial BH mass estimates
from Shen et al.\ 2011). Since the SDSS quasars reside in halos that are
typically a factor of a few less massive than CMASS galaxy hosts, either the
quasar BH mass in these lower-mass galaxies is over-massive compared with the
prediction from the local $M_{\rm BH}-M_{\rm bulge}$ relation, or the virial
mass estimates for SDSS quasars are systematically overestimated \citep[for
the latter possibility, see,
e.g.,][]{Shen_etal_2008b,Shen_Kelly_2010,Shen_Kelly_2012}. }

We now examine what constraints the luminosity-dependence of quasar bias at
$z\sim 0.5$ can place on cosmological quasar models. First, we derive a quick
constraint on the luminosity dependence of quasar bias by fitting a straight
line to the data. For simplicity we neglect (small) correlated errors among
these bias estimates due to the usage of the common galaxy sample in the
cross-correlation measurements. Using the four luminosity subsamples in the
two divisions, the slope constrained from the data is
\begin{eqnarray}
\frac{db_{Q}}{d\log L}&=&0.20\pm 0.34\quad \textrm{div 1} \\
&=&0.11\pm 0.32 \quad \textrm{div 2}\ ,
\end{eqnarray}
for $-23.5>M_i(z=2)>-25.5$. Thus the data are consistent with no luminosity
dependence over this luminosity range.

This weak luminosity dependence is in contrast to that of galaxy clustering
\citep[e.g.,][]{Norberg_etal_2001,Zehavi_etal_2005,Zehavi_etal_2011,
Coil_etal_2008,Coupon_etal_2012}. The SDSS main galaxy sample at $\langle
z\rangle\sim 0.1$ shows a strong positive luminosity dependence in galaxy
clustering \citep[][]{Zehavi_etal_2011}: $b_G(>L)\times
\sigma_8/0.8=1.06+0.21(L/L^*)^{1.12}$, where $L^*$ corresponds to
$M_r=-20.5$. For the $0.4<z<0.6$ galaxies in the Canada-France-Hawaii
Telescope Legacy Survey (CFHT-LS) sample \citep{Coupon_etal_2012},
$b_G(>L)=1.166+0.288(L/L^*)$ where $L^*$ corresponds to $M_g^*-5\log
h=-19.81$ (for all galaxies). The luminosity dependence of galaxy bias for
the CFHT-LS sample is shown in Fig.\ \ref{fig:b0_qso_gal_comp} and compared
to that of the quasar bias derived in this work. We have assumed that the
median quasar luminosity in our sample ($M_i(z=2)=-24.055$) corresponds to
the galaxy threshold luminosity with the same bias, which incidently
corresponds to a galaxy luminosity of $\approx L^*$. Based on this
comparison, a luminosity dependence of quasar clustering as strong as that
for galaxies is ruled out at the $\sim 95\%$ ($\sim 2\sigma$) confidence
level (CL). This result reflects a reasonably good correlation between galaxy
luminosity (and stellar mass) and halo mass, a correlation that appears to be
weaker between quasar luminosity and halo mass.


The linear bias for a population of quasars at fixed luminosity $L$ can be
expressed as \citep[e.g.,][]{Shen_2009}:
\begin{equation}
b_{Q}(L) = \int b_{h}(M_h)\frac{dP(M_h|L)}{dM_h}dM_h\ ,
\end{equation}
where $b_{h}(M_h)$ is the linear bias of halos with mass $M_h$, and
$dP(M_h|L)/dM_h$ is the distribution of host halo mass at fixed quasar
luminosity $L$. If we define an effective halo mass $\langle M_{h}\rangle(L)$
such that $b_h(\langle M_{h}\rangle)\equiv b_{Q}(L)$, the dependence of
$\langle M_{h}\rangle$ on $L$ determines the luminosity dependence of quasar
bias. As a toy model, we parameterize a relation $\langle
M_{h}\rangle(L)\propto L^\alpha$. A slope of $\alpha\approx 0.6\sim 0.75$ is
consistent with a model in which all quasars are shining at fixed Eddington
ratio, and their BH mass correlates with halo mass as $M_{\rm BH}\propto
M_h^{4/3-5/3}$ {\em with no scatter} (i.e., a ``light bulb'' model for
quasars). The scaling can be predicted from some analytical arguments
\citep[e.g.,][]{Silk_Rees_1998,Wyithe_Loeb_2003} or inferred from
observations of local dormant BHs
\citep[e.g.,][]{Ferrarese_2002,Baes_etal_2003} although scatter in the
relation is expected. Any scatter in the $M_{\rm BH}-M_h$ relation, and
dispersion in the Eddington ratio distribution, will lead to flattening in
the $\langle M_{h}\rangle-L$ correlation (i.e., reducing $\alpha$). Thus the
level of observed luminosity dependence of quasar bias places a constraint on
the scatter between halo mass and quasar luminosity for a given power-law
slope in the intrinsic correlation.

Fig. \ref{fig:b0_model_comp} (left) shows several realizations of this toy
model with different values of $\alpha$ in dotted lines. Models with large
$\alpha$ are less favorable compared with the data, although they cannot be
completely ruled out given the uncertainties in the measurements.

There are several more realistic, semi-analytical quasar models that can be
confronted with this observational constraint (see \S\ref{sec:intro} and
Appendix B of White et~al.\ 2012). It is beyond the scope of this paper to
compare these different models in detail or use our measurements to constrain
their model parameters \citep[cf.][]{Shankar_etal_2010,Shankar_etal_2010b}.


As a simple demonstration, we consider one semi-analytical quasar model from
\citet{Shen_2009}. This cosmological quasar model assumes that quasars are
triggered in halo major mergers, and adopts a quasar light curve model
composed of an Eddington-limited accretion phase and a power-law decaying
phase. This model can reproduce a variety of quasar observables, including
quasar clustering, luminosity function and Eddington ratio distributions over
a wide redshift range. In Fig.\ \ref{fig:b0_model_comp} (left) we show the
model predictions for the quasar bias as a function of luminosity at
$z=0.5-0.6$ as the gray shaded region. Although this model still predicts a
mild increase in quasar bias with luminosity, it matches the data very well.
The right panel of Fig.\ \ref{fig:b0_model_comp} displays the predicted
distribution of halo mass for quasars at several fixed luminosities. There is
considerable overlap in the range of halo masses for these quasar
luminosities, which dilutes the bias difference of these quasars with
different luminosities. The large dispersion in halo mass at fixed quasar
luminosity is caused by both the scatter between halo mass and BH mass (or
peak luminosity) and the luminosity evolution of individual quasars
\citep[see discussions in,
e.g.,][]{Lidz_etal_2006,White_etal_2008,Shen_2009,Shankar_etal_2010}.

We also compare the data with the prediction from a simple model connecting
halos and galaxies to quasars recently proposed by \citet{CW13}. This model
is a ``scattered light bulb'' model which assumes a linear relation between
galaxy mass and quasar BH mass, a lognormal distribution of quasar Eddington
ratios, and a constant duty cycle. The free parameters in this model are
tuned to match the observed quasar luminosity function over a wide redshift
range. The predicted luminosity dependence of quasar bias at $z=0.55$ from
their fiducial model (without satellite-hosted quasars) is shown as the blue
dashed line in the left panel of Fig.\ \ref{fig:b0_model_comp}. This model
predicts a luminosity dependence that is slightly stronger than that
predicted by the \citet{Shen_2009} model, although it is still consistent
with the data within $1\,\sigma$.  Inclusion of satellite hosted quasars
increases the predicted bias in the fainter bins by about 5\% while
negligibly changing the brighter bins. This marginally improves the agreement
with our data.

One might expect a stronger BH mass dependence of quasar clustering, because
the additional scatter between the instantaneous luminosity and BH mass
(i.e., the Eddington ratio distribution at fixed BH mass) has no effect here.
Quasar BH masses can be estimated with the virial BH mass estimators
\citep[e.g.,][]{Vestergaard_Peterson_2006}. We tested this hypothesis by
dividing the quasar sample using virial BH masses estimated in
\citet{Shen_etal_2011}, but did not find any significant dependence on virial
BH mass \citep[also see, e.g.,][]{Shen_etal_2009a}. This result, however,
could be due to the large statistical and systematic uncertainties of these
virial BH mass estimates \citep[e.g.,][]{Shen_etal_2008b}, or due to a large
scatter in the intrinsic correlation between halo mass and quasar BH mass.

\subsection{Halo occupation distribution modeling}\label{sec:disc2}

Next, we attempt to model our CCF measurements with simple Halo Occupation
Distribution (HOD) models \citep[for a review on halo models, see,
e.g.][]{Cooray_Sheth_2002}. This approach is an intuitive way to interpret
the observed CCF, and can offer insights on how galaxies and quasars form in
dark matter halos.

We fix the galaxy HOD by adopting parameters consistent with those in
\citet{White_etal_2011} from modeling the CMASS galaxy ACF, which reproduces
our DR10 CMASS ACF measurement. The large-scale galaxy bias parameter from
this set of HOD parameters is $b_G=2.00$. For the quasar HOD, we focus on two
types of parameterizations. Both types separate the contributions from
central and satellite quasars\footnote{In this work we use the term
``satellite quasar'' to refer to quasars hosted by satellite galaxies.} in
halos, and they differ in the form of the central quasar HOD. In the first
parameterization, the mean number of quasars located at the center of a halo
of virial mass $M$ is parameterized as
\begin{equation}
 \langle \Ncen(M)\rangle
= \frac{1}{2}\left[1+{\rm erf}\left(\frac{\log M-\log\Mmin}{\siglgM} \right)\right].
\end{equation}
This is a softened step function with characteristic mass scale $\Mmin$ and
transition width of $\siglgM$. We parameterize satellite quasars as a power
law with a low mass rolloff,
\begin{equation}
 \langle \Nsat(M)\rangle
= \exp\left(-\frac{M_0}{M}\right)\left(\frac{M}{M_1^\prime}\right)^\alpha .
\end{equation}
Such a quasar HOD parameterization is similar in form to the galaxy HOD
\citep[e.g.,][]{Zheng05,Zheng07}, and it is loosely motivated by cosmological
hydrodynamic simulation of AGN \citep[][]{DiMatteo08,Chatterjee_etal_2012}.
This five-parameter model ($\Mmin$, $\siglgM$, $M_0$, $M_1^\prime$, and
$\alpha$) has been applied to model the two-point auto-correlation functions
of $\langle z\rangle=1.4$ and $\langle z\rangle=3.2$ SDSS quasars
\citep[][]{Richardson12}. The second quasar HOD parameterization adopts the
same satellite HOD form, but it uses a log-normal form for the mean
occupation function of central quasars,
\begin{equation}
 \langle \Ncen(M)\rangle
= f_{\rm cen}\exp\left[-\frac{(\log M-\log M_{\rm cen})^2}{2\sigma_M^2} \right].
\end{equation}
This parameterization has 6 parameters in total (3 for satellite HOD and 3
for central HOD). Compared to the 5-parameter model, it reduces the number of
central quasars in massive halos. We will refer to the two types of HOD
parameterizations as 5-par and 6-par models, respectively.

For both parameterizations, we assume {\it no correlation between the
occupation numbers of central and satellite quasars and between galaxies and
quasars.} We also assume that the spatial distributions of both quasars and
galaxies inside halos follow the Navarro-Frenk-White (NFW) profile
\citep{Navarro97}. The variation and limitation of the quasar HOD
parameterizations will be discussed after presenting the main modeling
results.

\begin{figure*}[ht]
\plotone{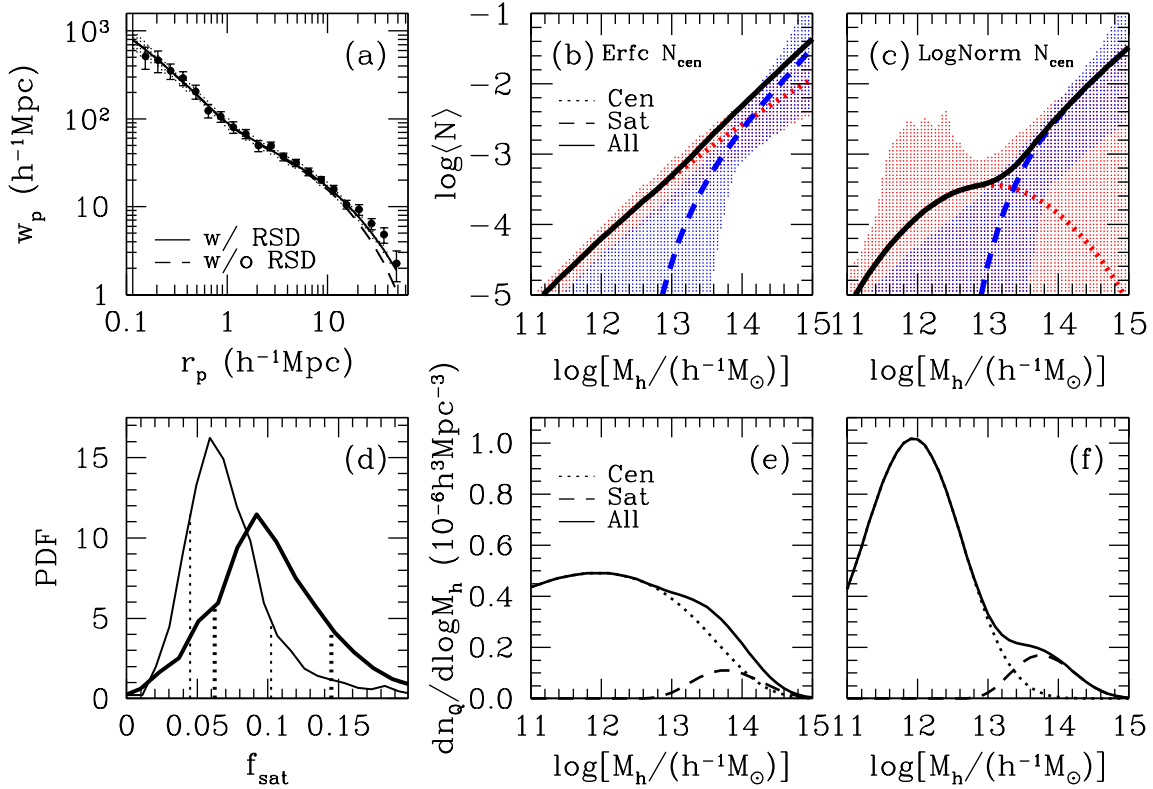} \caption{ \label{fig:hod} Results from HOD modeling
of the cross-correlation between galaxies and the full sample of quasars.
{\it Panel (a)}: HOD fit to the projected galaxy-quasar CCF. The solid curve
is the best-fit from the 5-par HOD model with the effect of residual redshift
space distortion (RSD) included. The shaded region is the envelope of the
fits from the 68.3\% of the models with the smallest $\chi^2$ values in the
MCMC chain. The dashed curve is the predicted $w_p$ with the above best-fit
HOD, if the effects of residual RSD were not included. {\it Panel (b)}: The
best-fit mean occupation function of quasars (solid) from the 5-par model,
decomposed into its central (dotted) and satellite (dashed) components. The
red and blue shaded regions are envelopes from the 68.3\% of models with the
lowest $\chi^2$ values for the central and satellite mean occupation
functions. {\it Panel (c)}: Same as {\it (b)}, but from the 6-par model. {\it
Panel (d)}: The fraction of satellite quasars in the full sample derived from
the HOD modeling. The thin and thick curves are from the 5-par and 6-par
models, respectively. Dotted lines enclose the central 68.3\% of each
distribution. {\it Panel (e)}: The contribution to the quasar number density
as a function of halo mass, decomposed into central (dotted) and satellite
(dashed) quasars, from the best-fit 5-par model. The curves are obtained from
the product of the mean occupation functions and the differential halo mass
function. The curves are also proportional to the probability distribution of
host halo mass of quasars. {\it Panel (f)}: Same as {\it (e)}, but from the
6-par model. See the text for details on the 5-par and 6-par models. }
\end{figure*}

The calculation of the galaxy-quasar two-point CCF in the HOD framework
follows similar procedures in \citet[][]{Zheng04}, \citet{Zehavi_etal_2005},
and \citet[][]{Tinker_etal_2005}. One improvement we have in the model is to
incorporate the effect of residual redshift-space distortion (RSD) when
computing the projected CCF from the real-space CCF, by applying the method
of \citet[][]{Kaiser87} to decompose the CCF into monopole, quadrupole, and
hexadecapole moments (also see \citealt{vandenBosch12}; J. Tinker, private
communication, 2009), which improves the modeling on large scales as we will
see later.

We model the cross-correlation between CMASS galaxies and the full sample of
quasars at the pair-weighted redshift $z=0.53$. We include the quasar number
density in calculating $\chi^2$, adopting a value of $2\times 10^{-6} h^3{\rm
Mpc}^{-3}$ with a 20\% fractional error (see Figure 2). A Markov Chain Monte
Carlo method is applied to probe the parameter space.

The main results from the HOD modeling are shown in Figure~\ref{fig:hod}. In
Figure~\ref{fig:hod}$(a)$, the solid curve is the best-fit $w_p$ from the
5-par model, with $\chi^2$/dof=26.6/18. 
The value of $\chi^2$ is about 1.4$\sigma$ higher than the expected mean
value 18, which is mostly contributed by the three points between
$20\hinvMpc$ and $40\hinvMpc$. While it is an acceptable fit, the slightly
higher $\chi^2$ may indicate that the model needs further improvement or that
the error bars and covariances on large scales are underestimated. The dashed
curve shows the predicted $w_p$ with the best-fit HOD if the residual RSD is
not included in the model. As expected, on scales much less than $\pi_{\rm
max}=70\hinvMpc$, the effect of residual RSD is small. However, on scales
close to $\pi_{\rm max}$, the effect starts to appear, e.g., about 40\% lower
in $w_p$ at $r_p\sim 50\hinvMpc$ if the residual RSD is neglected. The
$\chi^2$ from the $w_p$ with no RSD becomes $\chi^2$/dof=33.3/18, clearly
demonstrating that including the residual RSD does improve the fitting
significantly.

The best-fit mean occupation functions for the 5-par model are shown in
Figure~\ref{fig:hod}$(b)$, which can also be interpreted as the
mass-dependent duty cycle of the quasars in the full sample, i.e., the
fraction of halos hosting active quasars in the full sample. For central
quasars, a large transition width of the softened step function makes
$\langle \Ncen(M) \rangle$ behave like a power law with an index of $\sim
0.8$ above $10^{11}\hinvMsun$. Satellite quasars (with power law index $\sim
1.07$ in $\langle \Nsat\rangle$ at the high mass end) start to dominate
around $10^{14}\hinvMsun$. The overall occupation function resembles a power
law with index $\sim 0.95$.
The shaded regions delineate the envelopes from the first 68.3\% of the
models after sorting them in ascending order of $\chi^2$, which give us some
idea of the constraining power of the CCF on the quasar HOD. For central
quasars, the high-mass end is not well constrained -- the fast drop in halo
mass function toward the massive end makes quasars in massive halos
contribute little to the large scale bias and number density of quasars. For
satellite quasars, the constraints are tighter around the mass scale where
they become comparable in occupation number to the central quasars. This mass
scale also corresponds to the mass range of halos that have a significant
contribution to small-scale galaxy-quasar pairs. Other than this mass range,
the constraints on satellite HOD are loose.

Multiplying the best-fit mean occupation function with the differential halo
mass function, we obtain the contribution to the quasar number density from
halos of different masses, as shown in Figure~\ref{fig:hod}$(e)$. With
appropriate normalization, the curve also gives the probability distribution
of the host halo mass of the quasars in the full sample. While peaked around
$10^{12}\hinvMsun$, the host halos have a wide distribution in mass, about 4
dex in a full-width-half-maximum sense.
Marginalized over all models, the median host halo masses for central and
satellite quasars are $\log M_{\rm med,cen} = 11.60_{-0.39}^{+0.36}$ and
$\log M_{\rm med,sat} = 13.74_{-0.39}^{+0.27}$, respectively.


Figure~\ref{fig:hod}$(e)$ demonstrates that satellite quasars (dashed curve)
clearly make a non-negligible contribution to the full sample. The strong
small-scale clustering in the data requires the existence of satellite
quasars. Otherwise, the small-scale $w_p$ would become shallower. The
satellite fraction marginalized over all models is $f_{\rm
sat}=0.068_{-0.023}^{+0.034}$ (the thin curve in Figure~\ref{fig:hod}$(d)$).

With the adopted HOD parameterization, the 5-par model successfully
reproduces the observed galaxy-quasar CCF. The central quasar occupation
function appears to be a significantly softened step function
($\siglgM=2.73_{-0.21}^{+0.20}$).
Such a large transition width implies a large scatter in quasar luminosity at
any given halo mass. The large transition width also leads to a wide mass
range of host halos, which even extends to a few times $10^9\hinvMsun$,
a regime for dwarf galaxies. 
This result of low mass halos does not appear to be reasonable. Could it be
an artifact of the parameterization of the 5-par model? The $\langle
\Ncen(M)\rangle$ function is parameterized to be monotonically increasing
with mass towards an asymptotic value of unity (although it never reaches
unity in the mass range of interest). There are only two free parameters in
$\langle \Ncen(M)\rangle$, making a relatively tight connection between the
high-mass end and the low-mass end HOD. For example, while a higher $\langle
\Ncen(M)\rangle$ at the high mass end helps to reproduce the small-scale
clustering, it increases the large-scale bias, and as a response, the
occupation function must extend to low-mass halos to reduce the large scale
bias.

The 6-par model can explore the parameterization limitation, which allows the
high-mass occupation function of central quasars to cutoff exponentially. It
tends to mimic the lack of quasar activity in high mass halos where gas
accretion is likely suppressed. With this 6-par model, we find an almost
equally good fit to $w_p$, with $\chi^2$/dof=26.1/17, and the best-fit curve
is similar to that in Figure~\ref{fig:hod}$(a)$. The constraints on the mean
occupation functions (indicated by the shaded regions in
Figure~\ref{fig:hod}$(c)$) become less tight, especially for central quasars.
The host halo mass for central quasars now has a much narrower distribution
(see Figure~\ref{fig:hod}$(f)$), which is in a better agreement with the
prediction from the Shen (2009) model (See the right panel in Fig.\
\ref{fig:b0_model_comp}). Marginalized over all models, the median host halo
masses for central and satellite quasars are $\log M_{\rm med,cen} =
11.85_{-0.33}^{+0.25}$ and $\log M_{\rm med,sat} = 13.66_{-0.34}^{+0.26}$,
respectively. The satellite fraction from the 6-par model is $f_{\rm
sat}=0.099_{-0.036}^{+0.046}$ (see the thick curve in
Figure~\ref{fig:hod}$(d)$).



\begin{figure}[ht]
\plotone{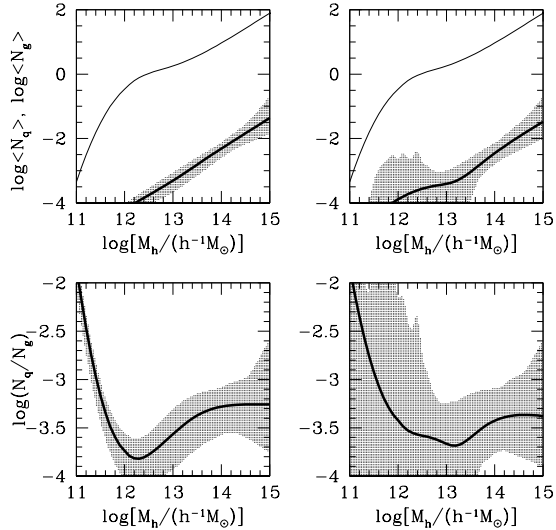} \caption{ \label{fig:nqng} {\em Top:} the mean (total)
occupation number of quasars and galaxies for the two quasar HOD
parameterization described in \S\ref{sec:disc2}. The galaxy HOD is the CMASS
HOD shifted to lower mass scales to mimic a $L>L^*$ galaxy sample, which
seems consistent with that in \citet{Coupon_etal_2012}, and roughly matches
the large-scale clustering of quasars. {\em Bottom:} the ratio between the
mean occupation numbers of quasars and galaxies. The shaded region indicates
the $68.3\%$ confidence range. For both quasar HOD parameterizations the
ratio of quasars to galaxies rises to a plateau at the high-mass end, but
the uncertainties are too large to confirm or rule out a decline in the
quasar fraction (per galaxy) in $>10^{14}\,M_\odot$ halos (e.g., clusters of
galaxies). }
\end{figure}

The high satellite fraction from either model is a somewhat surprising
result. With a similar 5-par parameterization, \citet{Richardson12} model the
2-point auto-correlation function of $0.5<z<2.5$ ($\bar{z}=1.4$) SDSS quasars
and infer a satellite fraction of $(7.4\pm 1.3) \times 10^{-4}$. Also from
HOD modeling of quasar clustering, \citet{Kayo_Oguri_2012} infer a satellite
fraction of $0.054_{-0.016}^{+0.017}$ for $0.6<z<2.2$ quasars. Although our
result is close to the latter one, the parameterizations are different
--- \citet{Kayo_Oguri_2012} assumes that both the central and satellite quasar
occupation functions have the same Gaussian form, differing only in the
amplitudes. The satellite fraction is mainly determined by the small-scale
clustering. In detail, for our quasar-galaxy CCF modeling, the result would
depend on the assumptions about the correlation between galaxies and quasars
inside halos and about the spatial distribution of satellite quasars and
galaxies inside halos. {\em This again highlights the ambiguity in HOD
parameterizations for the quasar population. }

One important distinction is that the quasar satellite fraction in our HOD
model is {\em not} the fraction of binary quasars (quasar pairs on 1-halo
scales). Many of the massive halos will only have one satellite quasar and no
central quasar, thus the actual binary quasar fraction would be substantially
lower than the satellite fraction. We still designate these quasars as
satellite quasars (even though they are the only quasar in the halo) because
they have a distinct intra-halo spatial distribution compared to central
quasars in our HOD modeling.

The clustering measurement can be well fit using different HOD
parameterization, as demonstrated by our 5-par and 6-par models. That is,
there exist large degeneracies in quasar HOD from the clustering data alone.
In addition to the 2-point correlation functions, we need other observables
(e.g., pairwise velocity distribution) to break the degeneracies and
constrain the connection between quasars and halos. We also need to rely on
theoretical work for a more physically motivated HOD parameterization to
model quasar clustering.

We also tried to model the HOD for our quasar luminosity subsamples, but the
constraints are poor given the increasingly larger measurement uncertainties.
Therefore we defer a more detailed HOD modeling of the luminosity dependence
of quasar clustering to future work with improved clustering measurements
(especially on small scales, see discussions in \S~4.4). The large-scale
quasar bias for the full sample from our HOD modeling is:
$b=1.27^{+0.08}_{-0.07}$ (5-par) and $b=1.26^{+0.08}_{-0.07}$ (6-par), which
are slightly lower, but consistent with our estimation in \S\ref{sec:cf}
within 1$\sigma$.

Finally, we comment on whether quasars are under-represented in massive halos
by examining the ratio of quasars to galaxies as a function of halo mass.
Fig.\ \ref{fig:nqng} shows the ratio of (central$+$satellite) quasars to
galaxies as a function of host halo mass, for the two HOD parameterizations
above. For the galaxy HOD we have simply shifted the CMASS HOD to lower mass
scales to approximate a $L>L^*$ galaxy sample, which seems to be consistent
with the results in \citet{Coupon_etal_2012}, and roughly matches the
large-scale clustering of quasars (see Fig.\ \ref{fig:b0_qso_gal_comp} and
caption thereof). The quasar-to-galaxy ratio rises to a plateau at high halo
masses in both HODs, but the uncertainties are large and we cannot confirm or
exclude a decline of quasar fraction (per galaxy) in $\gtrsim
10^{14}\,M_\odot$ halos (e.g., clusters of galaxies).

We tabulate the best-fit quasar HOD parameters and the adopted CMASS galaxy
HOD parameters in Table \ref{table:hod}, but we caution that the quasar HODs
are merely for future reference purposes and not for detailed physical
interpretation, given the large degeneracies discussed above.


\begin{table}
\centering
\caption{The adopted CMASS galaxy HOD parameters and the best-fit
parameters for the two quasar HOD parameterizations described in
\S\ref{sec:disc2}. All masses are in units of $h^{-1}M_\odot$. We caution
that the quasar HODs are merely for future reference purposes and not for
detailed physical interpretation, given the large degeneracies discussed in
\S\ref{sec:disc2}.} \scalebox{1.0}{
\begin{tabular}{lc | lc | lc}
\hline
\multicolumn{2}{c}{CMASS HOD} & \multicolumn{2}{c}{5-par quasar
HOD} & \multicolumn{2}{c}{6-par quasar HOD} \\
\multicolumn{2}{c}{Eqs. (8) and (9)} & \multicolumn{2}{c}{Eqs. (8) and (9)} &
\multicolumn{2}{c}{Eqs. (9) and (10)} \\
\hline
$\log M_{\rm min}$ & $13.14$ & $\log M_{\rm min}$ & $19.46^{+0.61}_{-0.64}$ & $\log M_{\rm cen}$ & $13.57^{+4.92}_{-1.41}$ \\
$\sigma_{\log M}$ & $0.485$ & $\sigma_{\log M}$ & $2.73^{+0.20}_{-0.21}$ & $\sigma_M$ & $0.91^{+0.82}_{-0.62}$ \\
$\log M_0$ & $13.01$ & $\log M_0$ & $12.74^{+0.86}_{-1.05}$ & $\log f_{\rm cen}$ & $-3.13^{+2.10}_{-0.46}$ \\
$\log M_1^{\prime}$ & $14.05$ & $\log M_1^{\prime}$ & $16.24^{+0.81}_{-0.51}$ & $\log M_0$ & $12.53^{+0.88}_{-1.02}$ \\
$\alpha$ & $0.97$ & $\alpha$ & $1.19^{+0.37}_{-0.33}$ & $\log M_1^{\prime}$ & $16.13^{+0.73}_{-0.40}$ \\
& & & & $\alpha$ & $1.21^{+0.29}_{-0.33}$\\
\hline
\end{tabular}}\label{table:hod}
\end{table}



\subsection{Mock catalog based interpretation}\label{sec:disc3}

\begin{figure*}
\includegraphics[width=0.95\textwidth]{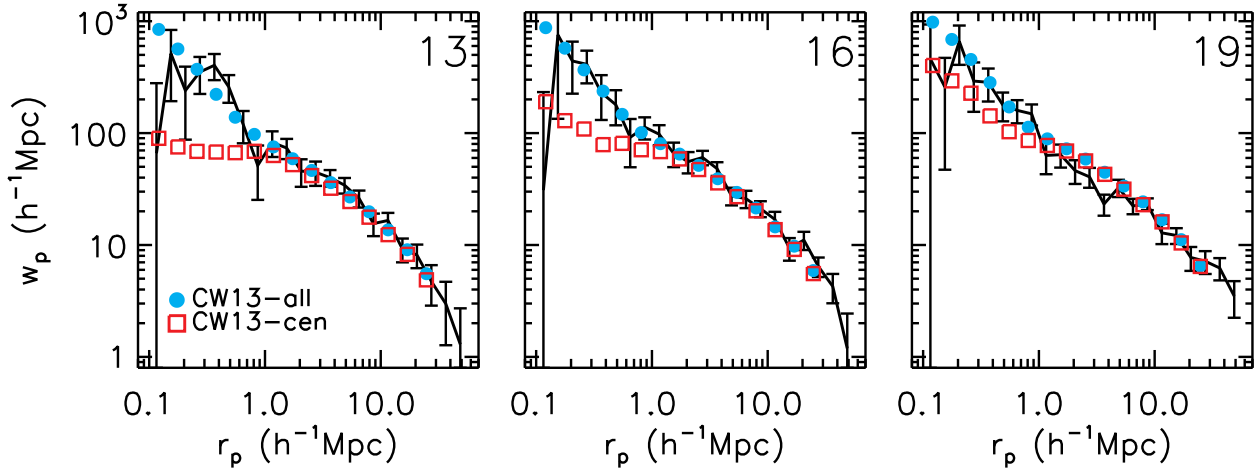}
\caption{Comparisons between the measured CCF and predictions from our mock
catalogs, for the three luminosity subsamples 13, 16 and 19 (see Table
\ref{tab:summary}). In each panel the black line with error bars is the
measured CCF, the red open squares are the prediction for mock quasar model
(1) and the cyan filled circles are the prediction for mock quasar model (2).
The errors on the predicted CCF are smaller than the observational errors,
and are suppressed for clarity. See text for details on the mock catalogs and
interpretations. } \label{fig:nbody_qso}
\end{figure*}

\begin{figure}
\includegraphics[width=0.9\linewidth]{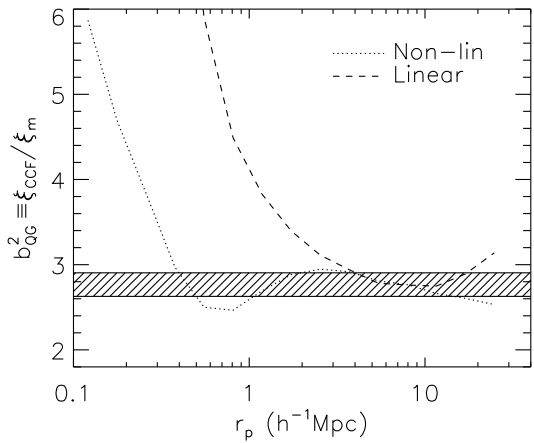}
\caption{Linear and non-linear biases of the CCF from one of our mock
catalogs. The underlying matter correlation function was computed using the
linear and non-linear power spectra from the simulation directly. The shaded
region encloses the $\pm 5\%$ range of the median non-linear bias within $r_p=4-16\,h^{-1}$Mpc.
Both the linear and non-linear biases show scale-dependence. The non-linear bias is computed
using the projected correlation function including redshift space distortions while the
linear bias calculation does not include redshift space distortions.
For scales $4<r_p<16\, {h^{-1}}$Mpc, the linear bias is roughly scale-independent.
This result motivated our choice of the fitting range in deriving the linear bias
in \S\ref{sec:cf}, for which the effects of scale-dependent bias and redshift
space distortions are negligible. } \label{fig:mock_bias}
\end{figure}

We now consider a mock catalog based approach to interpret the observed CCF
\citep[e.g.][]{Padmanabhan_etal_2009,White_etal_2011,CW13}. Compared with
analytic implementation of the HOD (\S\ref{sec:disc2}), the mock-based
approach directly uses simulated halo catalogs, thus avoiding using any
specific fitting formulae for the halo bias and abundance. Unfortunately it
can be subject to finite volume and finite resolution limitations. The basis
of our catalogs is a $2048^3$ particle N-body simulation of the $\Lambda$CDM
cosmology in a $700\,h^{-1}$Mpc box run with the {\sl TreePM\/} code
described in \citet{TreePM}.  This simulation has sufficient volume to probe
the CCF on the scales of relevance here while retaining sufficient force and
mass resolution to resolve the halos hosting CMASS galaxies and quasars.

We can populate the halos in the simulation using different models for the
relevant objects. The CMASS galaxies are placed in the halos using a HOD
similar to that described in \S\ref{sec:disc2}. The parameters are adjusted
to fit the small-scale clustering measured in \citet{White_etal_2011} and the
large-scale clustering measured in \citet{Anderson_etal_2012} for CMASS
galaxies. Since our purposes are primarily illustrative, we simply chose one
model which provides a good fit without attempting to propagate the
uncertainty in this model. This best-fit model is a very good fit to the
data. For the quasars we chose two different models based on the framework in
\citet[][CW13 for short]{CW13}. The CW13 framework assumes there is a linear
relation between galaxy stellar mass and BH mass with a scatter, and that the
BH shines as a quasar with a constant duty cycle, with its luminosity drawn
from a lognormal distribution with a constant mean Eddington ratio. This
simple model can reproduce the quasar luminosity function and large-scale
quasar bias for a wide range of redshifts.

For both quasar models we consider the cross-correlation on both large-
and small-scales is independent of the overall duty cycle of the quasars ---
a random dilution of the sample returns the same clustering on average.
The first model assumes quasars live at the centers of dark matter halos with
the quasar luminosity set by the stellar mass of the galaxy most likely to be
hosted by such a halo \citep[as in][]{CW13}. In the second model, quasars
live in both central and satellite galaxies, with the quasar luminosity set
by the stellar mass of the galaxy \citep[as in][]{CW13}. Comparison between
the two models shows the impact of quasars populating satellite galaxies.

Fig.~\ref{fig:nbody_qso} shows the CCF comparisons of our mock predictions
with the data, for the three luminosity subsamples: 13, 16 and 19 in Division
2 (see Table \ref{tab:summary}). In each panel, the black line with error
bars is the measured CCF, and the red (CW13-cen) and cyan (CW13-all) points
are our mock predictions for quasar model (1) and (2), respectively. Model
(1) where quasars only populate central galaxies does not provide a good
match to the small-scale CCF. On the other hand, Model (2) where quasars
populate both central and satellite galaxies provides a good match to the
overall CCF for three luminosity subsamples (although the model may
over-predict the CCF a little on scales of a few $h^{-1}$Mpc for sample 19).
The reason that the predicted CCF does not vary much over the three quasar
luminosity bins is that there is substantial overlap in the host halo mass
range for quasars in the three bins, due to the significant scatter between
host galaxy stellar mass and instantaneous quasar luminosity in the CW13
model ($\sim 0.4\,$dex). Since in Model (2), quasars are randomly subsampled
from galaxies regardless of their positions (with scatter), the overall
satellite fraction of quasars is roughly the same as for galaxies, i.e.,
$f_{\rm sat}\sim 10\%$ for the three luminosity samples shown in
Fig.~\ref{fig:nbody_qso}.
This satellite fraction is similar to that inferred from the 6-par HOD model
discussed in \S\ref{sec:disc2}. In reality, the situation may be more
complicated such that central galaxies might be less likely to host a quasar
than satellite galaxies in the most massive halos (e.g., clusters), which
will lead to changes in the satellite fraction. In addition, just as for our
HOD modeling, any enhanced probability of finding close galaxy-quasar pairs
(e.g., if quasars are triggered during interactions with companion galaxies)
will change our mock interpretation (which assumes galaxies and quasars are
statistically independent when populating the halos). Additional observations
of quasars in groups and clusters are required to probe these possibilities.

For our mocks, the mean quasar occupation number and the distribution of host
halo mass differ in detail from our best-fit HOD models in \S\ref{sec:disc2},
which again highlights the fact that there is a broad range of HOD parameter
space that can accommodate the observed CCF.

A side product of our mock-based modeling is a prediction for the
scale-dependence of the bias for the CCF. In Fig.\ \ref{fig:mock_bias} we
show the ratios of the CCF of our mock catalogs to the auto-correlation
function of the underlying dark matter computed from the linear and
non-linear matter power spectra from our simulation. The linear bias is
approximately constant over scales $\sim 4-16\,h^{-1}$Mpc. It is on the basis
of this modeling that we have chosen the fitting range quoted in
\S\ref{sec:cf}.

\subsection{The future}\label{sec:disc4}

Given the weak luminosity dependence of quasar clustering, one must
considerably improve the errors on the measurements to firm up a detection.
In addition, it is desirable to have a larger lever arm in quasar luminosity,
since the change in quasar linear bias with luminosity is slow.
With the cross-correlation technique the galaxy sample limits us to a fixed
area of sky. To go brighter we need to work at the highest redshift available
(both because of volume effects and because of the $z$-dependence of the
luminosity function). To go fainter we need to probe to dimmer objects in the
same area of sky. 


A major discriminant between quasars models lies in the less luminous quasars
(below $L_\star$). In older, or more simplified, models these quasars arise
from low-mass black holes accreting at close to the Eddington rate, whereas
in most modern models a significant fraction of them arise from higher mass
black holes accreting at a lower rate (and the prevalence of low accretion
rate black holes is particularly pronounced in the redshift range of interest
here).

Unlike most galaxy clustering measurements (especially those from SDSS),
quasar clustering measurements are still limited by statistical errors. Our
current cross-correlation sample only includes $\sim 2/3$ of the final CMASS
galaxy-DR7 quasar overlap sample. Thus we expect some improvement in the
clustering measurements using the final data release of BOSS. The
signal-to-noise ratio for Poisson noise dominated regimes (e.g., at small
scales) will increase by a factor of $\sim 1.2$. For large-scale bins where
errors are correlated, we expect improvements somewhat smaller than this. In
any case, the final cross-correlation sample will have a more uniform sky
coverage than the current sample, which may eliminate some systematic
problems.


\section{conclusions}\label{sec:con}

In this paper we presented the cross-correlation function measurements
between quasars and galaxies at $z\sim 0.5$ using a spectroscopic quasar
sample from SDSS DR7 and a BOSS CMASS galaxy sample from SDSS-III DR10. Our
cross-correlation sample contains 8,198 quasars and 349,608 BOSS (CMASS)
galaxies. Our main results are the following:

\begin{itemize}

\item The CCF can be well described by a power-law model
    $\xi_{QG}=(r/r_0)^{-\gamma}$ for scales $r_p=[2,25]\,h^{-1}$Mpc with
    $r_0=6.61\pm0.25\,h^{-1}$Mpc and $\gamma=1.69\pm0.07$. The
    large-scale quasar linear bias is estimated to be $b_Q=1.38\pm0.10$
    at $\langle z\rangle\sim 0.53$. This bias infers that quasars at
    these redshift reside in halos with typical mass of $\sim 4\times
    10^{12}\,h^{-1}M_\odot$ (using the Tinker et~al. 2005 fitting
    formula), similar to quasar clustering measurements at high-redshift,
    but lower than the typical halo mass $\sim 10^{13}\,h^{-1}M_\odot$
    for massive galaxies in SDSS. Thus most of these low-redshift quasars
    are not the descendants of their high-redshift counterparts, which
    would have evolved into more massive and more biased systems (such as
    the hosts of CMASS galaxies).

\item We found weak luminosity dependence of the large-scale quasar
    linear bias, over the luminosity range $-23.5>M_i(z=2)>-25.5$ probed
    by our sample. This result is generally consistent with other quasar
    clustering measurements at different redshifts. This weak luminosity
    dependence suggests that quasars with fixed luminosity spread over a
    broad range of host halo masses, in qualitative and quantitative
    agreement with predictions from several theoretical models
    \citep[e.g.,][]{Lidz_etal_2006,Shen_2009,CW13}.

\item We performed HOD and mock catalog-based modeling of the measured
    CCF. For the HOD modeling, we found large degeneracies in the HOD
    parameterizations such that different HODs can reproduce the CCF
    equally well, with different host halo mass distributions and
    satellite fractions. This result highlights the limitations and
    ambiguities in the standard HOD approach for modeling the quasar
    population. Additional information is needed in order to break the
    degeneracies in the quasar HOD models.

    For the mock-based approach, we found the simple model in
    \citet{CW13} that relates quasars to galaxies can reproduce the CCF
    reasonably well. Under such a model framework, we need a satellite
    fraction of quasars (i.e., fraction of quasars hosted by satellite
    galaxies) of $f_{\rm sat}\sim 10\%$. Just as for the HOD-based
    modeling, however, we cannot rule out other models by which quasars
    can inhabit dark matter halos and produce the same CCF.

    The difficulty of finding a unique HOD model for quasars probably
    lies primarily in the fact that quasars are a sparse population with
    an unknown duty cycle relative to halos (or galaxies). The large
    scatter between quasar luminosity and halo mass also makes it
    difficult to use luminosity-dependent clustering as an additional
    constraint in quasar HOD modeling.


\end{itemize}

With the upcoming data release of the BOSS survey, we will eventually have a
spectroscopic CCF sample with $\sim 50\%$ more quasars and more CMASS
galaxies with the final SDSS-III data release. The new data will increase the
cross-pair counts by $\sim 50\%$. On small scales ($r_p\lesssim 1\,h^{-1}{\rm
Mpc}$) where Poisson statistics dominate, we therefore expect $\sim 20\%$
improvement in the errors of $w_p$ measurements. These changes will
potentially be able to reveal differences in the small-scale clustering when
binned in quasar luminosity. In the short term, we also plan to measure the
CCF using spectroscopic SDSS-DR7 quasars and the photometric CMASS galaxy
sample, which will have the same cross-sample coverage as the final
spectroscopic CMASS sample and is free of fiber collision losses. Future
deeper galaxy and quasar surveys over large areas can improve the pair
statistics further, and at the same time increase the
dynamical range in quasar luminosity.

\acknowledgements

YS acknowledges support from the Smithsonian Astrophysical Observatory (SAO)
through a Clay Postdoctoral Fellowship and from Carnegie Observatories
through a Hubble Fellowship from Space Telescope Science Institute. Support
for Program number HST-HF-51314.01-A was provided by NASA through a Hubble
Fellowship grant from the Space Telescope Science Institute, which is
operated by the Association of Universities for Research in Astronomy,
Incorporated, under NASA contract NAS5-26555. ZZ and IZ acknowledge partial
support by NSF grant AST-0907947.

Funding for SDSS-III has been provided by the Alfred P. Sloan Foundation, the
Participating Institutions, the National Science Foundation, and the U.S.
Department of Energy Office of Science. The SDSS-III web site is
http://www.sdss3.org/.

SDSS-III is managed by the Astrophysical Research Consortium for the
Participating Institutions of the SDSS-III Collaboration including the
University of Arizona, the Brazilian Participation Group, Brookhaven National
Laboratory, University of Cambridge, Carnegie Mellon University, University
of Florida, the French Participation Group, the German Participation Group,
Harvard University, the Instituto de Astrofisica de Canarias, the Michigan
State/Notre Dame/JINA Participation Group, Johns Hopkins University, Lawrence
Berkeley National Laboratory, Max Planck Institute for Astrophysics, Max
Planck Institute for Extraterrestrial Physics, New Mexico State University,
New York University, Ohio State University, Pennsylvania State University,
University of Portsmouth, Princeton University, the Spanish Participation
Group, University of Tokyo, University of Utah, Vanderbilt University,
University of Virginia, University of Washington, and Yale University.

Funding for the SDSS and SDSS-II has been provided by the Alfred P. Sloan
Foundation, the Participating Institutions, the National Science Foundation,
the U.S. Department of Energy, the National Aeronautics and Space
Administration, the Japanese Monbukagakusho, the Max Planck Society, and the
Higher Education Funding Council for England. The SDSS Web Site is
http://www.sdss.org/.

The SDSS is managed by the Astrophysical Research Consortium for the
Participating Institutions. The Participating Institutions are the American
Museum of Natural History, Astrophysical Institute Potsdam, University of
Basel, University of Cambridge, Case Western Reserve University, University
of Chicago, Drexel University, Fermilab, the Institute for Advanced Study,
the Japan Participation Group, Johns Hopkins University, the Joint Institute
for Nuclear Astrophysics, the Kavli Institute for Particle Astrophysics and
Cosmology, the Korean Scientist Group, the Chinese Academy of Sciences
(LAMOST), Los Alamos National Laboratory, the Max-Planck-Institute for
Astronomy (MPIA), the Max-Planck-Institute for Astrophysics (MPA), New Mexico
State University, Ohio State University, University of Pittsburgh, University
of Portsmouth, Princeton University, the United States Naval Observatory, and
the University of Washington.

\begin{appendix}

%
%

\begin{figure}
 \centering
 \includegraphics[width=0.9\linewidth]{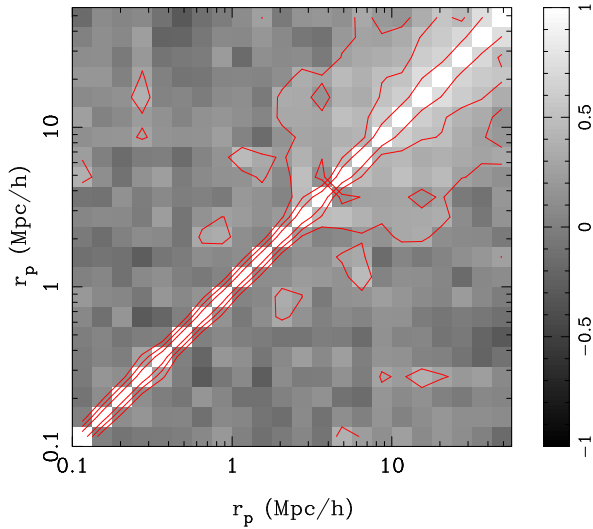}
 \caption{
   Correlation matrix of $w_p(r_p)$ for the full sample cross correlation
   (DR10 CMASS galaxies with DR7 uniform quasars). This is the normalized
   covariance matrix, i.e.~correlation matrix, such that the diagonal
   elements are unity calculated on 50 jackknife samples.
   The contours correspond to values of $0.75$, $0.50$, and $0.25$.
 }
 \label{fig:full_cov}
\end{figure}

We estimate errors on our clustering measurements using the jackknife
resampling technique (as discussed in Section~\ref{sec:cf}).  We use the full
covariance, which includes the correlation between bins in the correlation
function as shown in Figure~\ref{fig:full_cov}.  We use a fiducial value of
$50$ jackknife regions, which we define such that each region has the same
unmasked area on the sky and is roughly rectangular (where possible). In this
section, we evaluate some of the effects on the errors due to varying the
number of jackknife regions for our measurements of the projected two-point
correlation function. Specifically, we compare error estimates on our cross
correlation measurement for the full sample using 10, 25, 50, 75, and 100
jackknife regions.

\begin{figure*}
 \centering
 \includegraphics[width=0.9\linewidth]{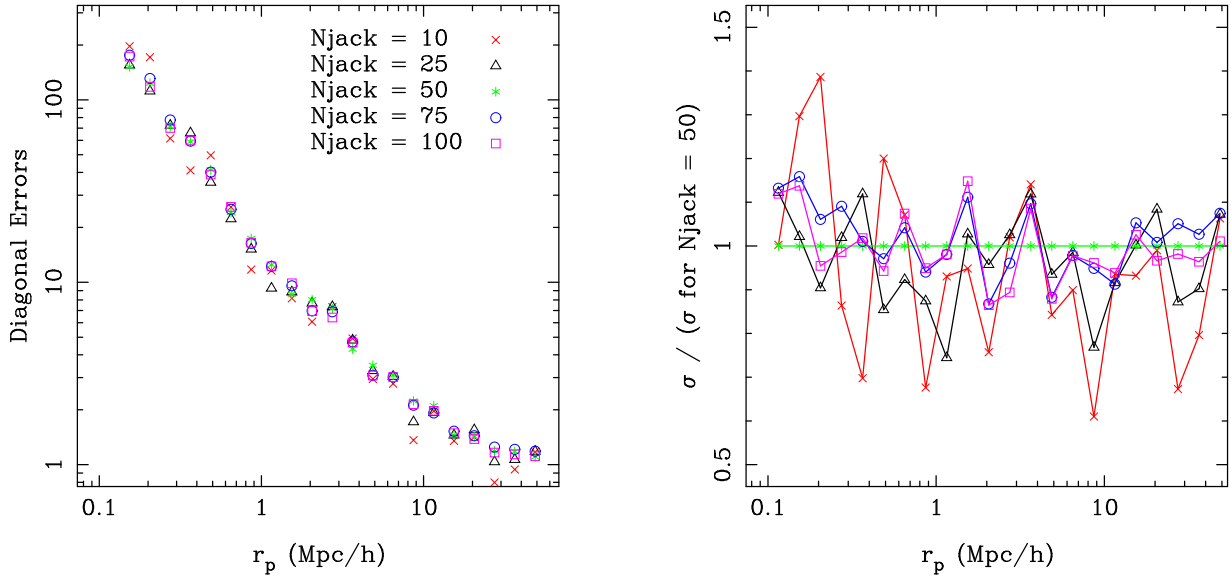}
 \caption{{\em Left:} the 1$\sigma$ (diagonal) errors calculated from varying number of jackknife
samples: 10, 25, 50, 75, 100. Our fiducial choice is 50 jackknife samples.
{\em Right:} the ratio of the diagonal errors with different numbers of
jackknife samples to those using 50 jackknife samples.
 }
 \label{fig:diag}
\end{figure*}

\begin{figure*}
 \centering
 \includegraphics[width=0.9\linewidth]{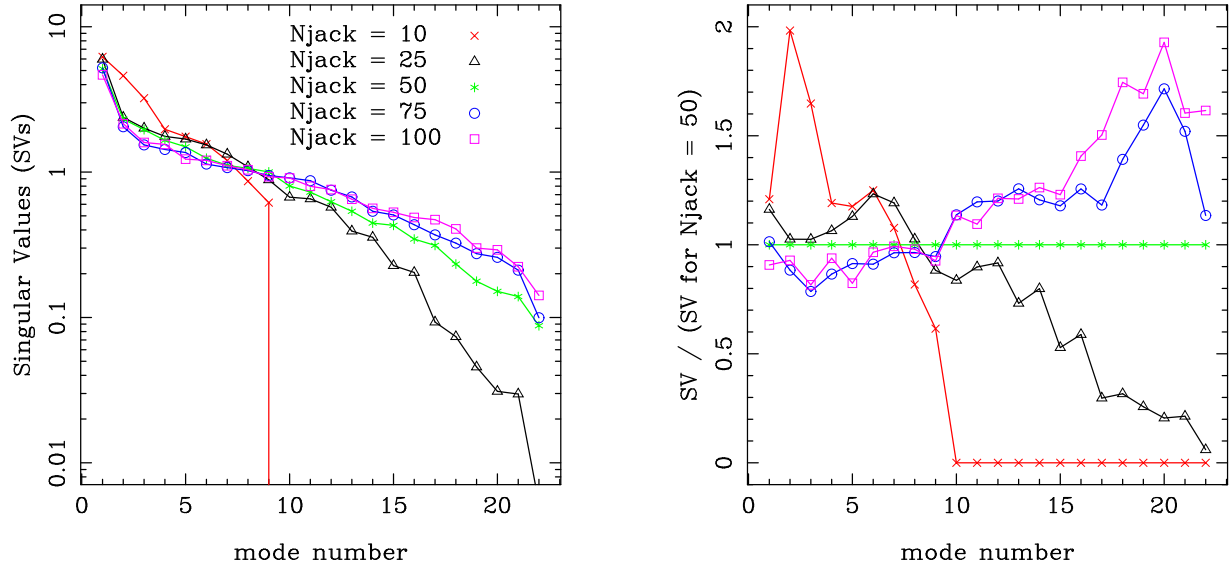}
 \caption{
   Singular values (or eigenvalues) obtained by performing a singular value decomposition
   (SVD) on the correlation matrix estimated from the different number of jackknife
   samples.
 }
 \label{fig:sv}
\end{figure*}

\begin{table*}\label{tab:CM}
\caption{Normalized covariance matrices from the full CCF sample and from
individual CCF subsamples. The corresponding diagonal errors are tabulated in
Table \ref{table:wp}. A portion (i.e., the covariance matrix from the full
CCF sample) is shown here for its content. The table is available in its
entirety in the electronic version of this paper.}
\scalebox{0.65}{
\begin{tabular}{l|rrrrrrrrrrrrrrrrrrrrrr}
\hline
$r_p\ [h^{-1}{\rm Mpc}]$ & $ 0.115$ & $ 0.154$ & $ 0.205$ & $
0.274$ & $ 0.365$ & $ 0.487$ & $ 0.649$ & $ 0.866$ & $ 1.155$ & $ 1.540$ & $
2.054$ & $ 2.738$ & $ 3.652$ & $ 4.870$ & $ 6.494$ & $ 8.660$ & $11.548$ &
$15.399$ & $20.535$ & $27.384$ & $36.517$ & $48.697$ \\
\hline
 0.115 & $ 1.000$ & $-0.002$ & $-0.087$ & $-0.047$ & $ 0.001$ & $ 0.137$ & $ 0.015$ & $ 0.141$ & $-0.019$ & $-0.023$ & $-0.161$ & $-0.035$ & $-0.083$
 & $ 0.273$ & $ 0.305$ & $ 0.049$ & $ 0.114$ & $-0.086$ & $ 0.078$ & $ 0.123$ & $ 0.022$ & $-0.182$ \\
 0.154 &               & $ 1.000$ & $ 0.082$ & $ 0.106$ & $ 0.028$ & $-0.010$ & $-0.179$ & $ 0.016$ & $ 0.058$ & $-0.029$ & $-0.116$ & $ 0.104$ &
 $-0.215$ & $ 0.000$ & $-0.077$ & $ 0.197$ & $ 0.008$ & $ 0.147$ & $ 0.125$ & $ 0.147$ & $ 0.107$ & $ 0.136$ \\
 0.205 &               &               & $ 1.000$ & $ 0.055$ & $-0.097$ & $-0.042$ & $ 0.230$ & $ 0.126$ & $-0.164$ & $-0.127$ & $-0.001$ & $ 0.101$
 &
 $ 0.057$ & $-0.155$ & $-0.220$ & $ 0.013$ & $-0.015$ & $ 0.109$ & $-0.070$ & $-0.069$ & $ 0.091$ & $ 0.052$ \\
 0.274 &               &               &               & $ 1.000$ & $ 0.226$ & $ 0.058$ & $-0.375$ & $-0.055$ & $ 0.074$ & $-0.213$ & $ 0.021$ &
 $-0.078$ & $-0.050$ & $ 0.034$ & $-0.061$ & $ 0.229$ & $ 0.229$ & $ 0.389$ & $ 0.262$ & $ 0.248$ & $ 0.213$ & $ 0.087$ \\
 0.365 &               &               &               &               & $ 1.000$ & $ 0.071$ & $-0.118$ & $ 0.340$ & $ 0.220$ & $-0.050$ & $ 0.138$ &
 $ 0.284$ & $ 0.034$ & $ 0.110$ & $-0.006$ & $ 0.068$ & $ 0.089$ & $ 0.144$ & $ 0.002$ & $-0.067$ & $-0.259$ & $ 0.030$ \\
 0.487 &               &               &               &               &               & $ 1.000$ & $ 0.027$ & $-0.049$ & $-0.061$ & $ 0.091$ &
 $-0.068$ & $-0.172$ & $-0.072$ & $-0.077$ & $-0.253$ & $ 0.047$ & $-0.038$ & $ 0.041$ & $-0.190$ & $-0.190$ & $-0.214$ & $-0.302$ \\
 0.649 &               &               &               &               &               &               & $ 1.000$ & $ 0.154$ & $-0.070$ & $ 0.010$ &
 $
 0.309$ & $ 0.133$ & $ 0.019$ & $-0.007$ & $-0.087$ & $ 0.098$ & $ 0.053$ & $-0.094$ & $-0.068$ & $-0.172$ & $-0.189$ & $-0.210$ \\
 0.866 &               &               &               &               &               &               &               & $ 1.000$ & $ 0.024$ & $
 0.033$ & $ 0.266$ & $ 0.269$ & $-0.108$ & $ 0.144$ & $ 0.150$ & $ 0.147$ & $ 0.051$ & $ 0.092$ & $ 0.029$ & $ 0.158$ & $-0.088$ & $-0.068$ \\
 1.155 &               &               &               &               &               &               &               &               & $ 1.000$ &
 $-0.087$ & $ 0.140$ & $ 0.180$ & $ 0.070$ & $ 0.170$ & $ 0.389$ & $ 0.123$ & $ 0.124$ & $ 0.194$ & $ 0.082$ & $-0.064$ & $-0.022$ & $ 0.133$ \\
 1.540 &               &               &               &               &               &               &               &               &
 & $ 1.000$ & $-0.087$ & $ 0.031$ & $ 0.105$ & $ 0.282$ & $ 0.276$ & $-0.060$ & $ 0.084$ & $ 0.020$ & $ 0.190$ & $ 0.333$ & $ 0.233$ & $ 0.157$ \\
 2.054 &               &               &               &               &               &               &               &               &
 &               & $ 1.000$ & $ 0.231$ & $ 0.047$ & $ 0.091$ & $ 0.097$ & $ 0.076$ & $ 0.280$ & $ 0.245$ & $ 0.127$ & $ 0.072$ & $-0.137$ & $-0.165$
 \\
 2.738 &               &               &               &               &               &               &               &               &
 &               &               & $ 1.000$ & $ 0.389$ & $ 0.350$ & $ 0.166$ & $ 0.297$ & $ 0.153$ & $ 0.336$ & $ 0.285$ & $ 0.077$ & $-0.149$ & $
 0.111$ \\
 3.652 &               &               &               &               &               &               &               &               &
 &               &               &               & $ 1.000$ & $ 0.110$ & $ 0.269$ & $ 0.173$ & $ 0.175$ & $ 0.216$ & $ 0.262$ & $ 0.063$ & $-0.011$ &
 $ 0.208$ \\
 4.870 &               &               &               &               &               &               &               &               &
 &               &               &               &               & $ 1.000$ & $ 0.487$ & $ 0.429$ & $ 0.268$ & $ 0.245$ & $ 0.355$ & $ 0.270$ & $
 0.119$ & $ 0.004$ \\
 6.494 &               &               &               &               &               &               &               &               &
 &               &               &               &               &               & $ 1.000$ & $ 0.363$ & $ 0.346$ & $ 0.311$ & $ 0.407$ & $ 0.423$ &
 $
 0.275$ & $ 0.236$ \\
 8.660 &               &               &               &               &               &               &               &               &
 &               &               &               &               &               &               & $ 1.000$ & $ 0.496$ & $ 0.598$ & $ 0.547$ & $
 0.294$ & $ 0.224$ & $ 0.207$ \\
11.548 &               &               &               &               &               &               &               &               &
&               &               &               &               &               &               &               & $ 1.000$ & $ 0.601$ & $ 0.565$ & $
0.425$ & $ 0.327$ & $ 0.181$ \\
15.399 &               &               &               &               &               &               &               &               &
&               &               &               &               &               &               &               &               & $ 1.000$ & $ 0.703$
& $ 0.548$ & $ 0.440$ & $ 0.385$ \\
20.535 &               &               &               &               &               &               &               &               &
&               &               &               &               &               &               &               &               &               & $
1.000$ & $ 0.679$ & $ 0.488$ & $ 0.482$ \\
27.384 &               &               &               &               &               &               &               &               &
&               &               &               &               &               &               &               &               &               &
& $ 1.000$ & $ 0.649$ & $ 0.403$ \\
36.517 &               &               &               &               &               &               &               &               &
&               &               &               &               &               &               &               &               &               &
&               & $ 1.000$ & $ 0.694$ \\
48.697 &               &               &               &               &               &               &               &               &
&               &               &               &               &               &               &               &               &               &
&               &               & $ 1.000$ \\
\hline
\end{tabular}}
\end{table*}

The number of jackknife regions is somewhat arbitrary \citep[see detailed
discussion in][]{Norberg_etal_2009}. Using too few jackknife samples will
result in a low number of realizations to estimate the variance, and can
formally cause the covariance matrix to become singular (when the number of
samples is less than the number of bins). The use of too many jackknife
regions causes each region to become small in area (and therefore volume) and
can inaccurately represent the cosmic (sample) variance in the large-scale
errors. At a minimum, we must ensure the size of each jackknife region is
significantly larger than the largest scales we measure in the data.

We first investigate the magnitude of the diagonal errors, which we show in
Figure~\ref{fig:diag}. The values of $\sigma$ can vary by up to $30-40$\%,
but are otherwise roughly equivalent. There is no systematic bias in the
values that affects one choice more than any other across all the bins. A
lower number of jackknife samples, however, results in a larger variation in
the values, as we would expect.

To quantify how well we resolve the structure of the correlation matrix (e.g.
Figure~\ref{fig:full_cov}), we perform a singular value decomposition (SVD)
on the correlation matrix. The SVD effectively rotates the matrix into an
orthogonal space which can be thought of as a combination of eigenvectors and
eigenvalues. The singular values (SVs) are eigenvalues (defined to be
positive) which are the multiplicative amplitude of the corresponding
(normalized) eigenvector. The SVs are typically numbered such that they are
monotonically decreasing, and can be interpreted as a measure of the
``importance'' of each mode in terms of contributing to the observed
structure in the full correlation matrix. For example, an $N$ by $N$ diagonal
correlation matrix (i.e. the identity matrix) would result in $N$ SVs that
were all equal in value. The ratio of the largest SV divided by the smallest
SV is referred to as the condition number, and if significantly large can
result in poor numerical results when the matrix is inverted (i.e. an
ill-conditioned matrix) which is performed in model fitting.

We show the SVs for our correlation matrices in Figure~\ref{fig:sv}. We
clearly see our expectation of the ill-conditioned matrix for $N_{\rm jack} =
10$ since we are using 22 bins. A larger $N_{\rm jack}$ results in a better
conditioned matrix (a line that appears more flat as the SVs vary less). We
also notice quickly diminishing returns for larger numbers of samples: while
there is a dramatic difference between $10$ and $50$ samples, it is much less
of a difference for the larger numbers of jacknife samples.

We conclude from these investigations that using less than $50$ jackknife
samples could be troubling. Taking into account the area coverage of our data
($4122$ deg$^2$), $50$ jackknife samples result in each jackknife region
covering about $82$ deg$^2$ (roughly $9$ deg or less on a side). As our
statistical errors are significantly larger than the galaxy autocorrelation
function \citep[e.g. ][]{Zehavi_etal_2011}, we are not overly concerned about
resolving each element of the covariance matrix. Our choice of $50$ jackknife
samples is a factor of 2 larger than the number of bins.

\end{appendix}

\bibliographystyle{apj}


\end{document}